\documentclass[acmsmall]{acmart}

\usepackage{booktabs} 
\usepackage{enumitem}
\usepackage{makecell}
\usepackage{textcomp}
\usepackage{amsmath}
\usepackage{algorithm}
\usepackage[noend]{algpseudocode}
\usepackage{pdfpages}
\usepackage{graphicx}

\usepackage{amssymb}
\usepackage{mathrsfs}
\usepackage{xcolor}
\usepackage{epsfig}
\usepackage{multirow}
\usepackage{enumerate}
\usepackage{bm}
\usepackage{subfigure}
\usepackage{url}
\usepackage{footmisc}
\usepackage{lineno}

\DeclareMathOperator*{\argmax}{arg\,max}

\def \bu {\mathbf{u}}

\def \bx {\mathbf{x}}
\def \bbE {\mathbb{E}}
\def \mt {\mathsf{T}}

\def \cP {\mathcal{P}}
\def \cN {\mathcal{N}}
\def \bB {\mathbf{B}}
\def \bx {\mathbf{x}}
\def \bu {\mathbf{u}}
\def \bp {\mathbf{p}}
\def \bf {\mathbf{f}}

\AtBeginDocument{%
  \providecommand\BibTeX{{%
    \normalfont B\kern-0.5em{\scshape i\kern-0.25em b}\kern-0.8em\TeX}}}

\setcopyright{acmcopyright}
\acmJournal{TOIS}
\acmYear{2021} \acmVolume{39} \acmNumber{4} \acmArticle{40}
\acmMonth{8} \acmPrice{15.00}\acmDOI{10.1145/3446427}

\acmBooktitle{Woodstock '20: ACM Transactions on Information Systems,
  June 03--05, 2020, Woodstock, NY}
\acmPrice{15.00}
\acmISBN{978-1-4503-XXXX-X/20/06}

\begin{document}

\title{Seamlessly Unifying Attributes and Items: Conversational Recommendation for Cold-Start Users}

\author{Shijun Li}
\authornote{Both authors contributed equally to this research.}
\email{lishijun@mail.ustc.edu.cn}
\orcid{1234-5678-9012}
\affiliation{%
  \institution{University of Scinece and Technology of China}
  \streetaddress{443 Huangshan Road}
  \city{He Fei}
  \country{China}
  \postcode{230027}
}
\author{Wenqiang Lei}
\authornote{Wenqiang Lei is the corresponding author.}
\authornotemark[1]
\email{wenqianglei@gmail.com}
\affiliation{%
  \institution{National University of Singapore}
  \streetaddress{13 Computing Drive, National University of Singapore}
  \city{Singapore}
  \country{Republic of Singapore}}
  \postcode{117417}

\author{Qingyun Wu}
\email{qw2ky@virginia.edu}
\affiliation{%
  \institution{University of Virginia}
  \streetaddress{Computer Science 85 Engeers's way}
  \city{Charlottesville}
  \state{VA}
  \country{United States}
  \postcode{22903}
}

\author{Xiangnan He}
\email{hexn@mail.ustc.edu.cn}
\affiliation{%
 \institution{University of Scinece and Technology of China}
 \streetaddress{443 Huangshan Road}
 \city{He Fei}
  \country{China}
  \postcode{230027}
  }

\author{Peng Jiang}
\email{jp2006@139.com}
\affiliation{%
  \institution{Kuaishou Inc.}
  \streetaddress{6 West Shangdi Road }
  \city{Haidian District}
  \state{Beijing}
  \country{China}}
  \postcode{100085}
  

\author{Tat-Seng Chua}
\email{dcscts@nus.edu.sg}
\affiliation{%
  \institution{National University of Singapore}
  \streetaddress{13 Computing Drive, National University of Singapore}
  \city{Singapore}
  \country{Republic of Singapore}
  \postcode{117417}}

\renewcommand{\shortauthors}{Li and Lei, et al.}

\begin{abstract}
Static recommendation methods like collaborative filtering suffer from the inherent limitation of performing real-time personalization for cold-start users. 
Online recommendation, e.g., multi-armed bandit approach, addresses this limitation by interactively exploring user preference online and pursuing the exploration-exploitation (EE) trade-off. 
However, existing bandit-based methods model recommendation actions homogeneously. Specifically, they only consider the \textit{items} as the arms, being incapable of handling the \textit{item attributes}, which naturally provide interpretable information of user's current demands and can effectively filter out undesired items.  In this work, we consider the conversational recommendation for cold-start users, where a system can both ask the attributes from and recommend items to a user interactively.
This important scenario was studied in a recent work ~\cite{zhang2020toward}. However, it employs a hand-crafted function to decide when to ask attributes or make recommendations. Such separate modeling of attributes and items makes the effectiveness of the system highly rely on the choice of the hand-crafted function, thus introducing fragility to the system. To address this limitation, we seamlessly unify attributes and items in the same arm space and achieve their EE trade-offs automatically using the framework of Thompson Sampling. Our \textit{Conversational Thompson Sampling} (ConTS) model holistically solves all questions in conversational recommendation by choosing the arm with the maximal reward to play. Extensive experiments on three benchmark datasets show that ConTS outperforms the state-of-the-art methods \textit{Conversational UCB} (ConUCB)~\cite{zhang2020toward} and \textit{Estimation–Action–Reflection} model~\cite{lei20estimation} in both 
metrics of success rate and average number of conversation turns.
\end{abstract}



\maketitle

\section{introduction} \label{intro}

Recommendation system plays an increasingly important role in the current era of information explosion. In the past decades, most recommendation research has focused on advancing modeling historical user behaviours, proposing various methods like traditional collaborative filtering \cite{BPR, fastMF}, content- and context- aware filtering~\cite{ACF, FM}, to the recently prevalent neural network methods~\cite{NCF, cheng2016wide} and graph-based methods~\cite{wang2019neural, LightGCN}. We call such methods as \emph{static recommendation} as they learn static user preference from past interaction history, which inevitably requires an adequate amount of past behaviour histories for each user. However, the dependency of historical data makes such methods suffer from the cold-start scenario where new users come with no past historical data. 

Cold-start users typically interact with the system very shortly but expect to receive high-quality recommendations. Moreover, their preferences might dynamically change over time, making the recommendation even harder. The recent emerging application and technology of \textit{conversational recommendation system} (CRS) bring a promising solution to the cold-start problem. A CRS is envisioned as an interactive system that can ask the user preference towards item attributes before making the recommendations. With preferred attributes known, { the item pool can be significantly reduced~\cite{sun2018conversational}  } and the candidate items can be better scored~\cite{lei20estimation}, leading to more desired recommendations. 
Owing to this great potential, conversational recommendation becomes an emerging topic which attracts extensive research attention recently~\cite{christakopoulou2016towards, christakopoulou2018q,zhang2018towards,sun2018conversational, priyogi2019preference, yu2019visual, sardella2019approach, zhang2020toward,chen-etal-2019-towards}.

\begin{figure}
  \centering
  \subfigure[Taobao]{
    \label{fig:subfig:a} 
    \centering
    \includegraphics[width=2in]{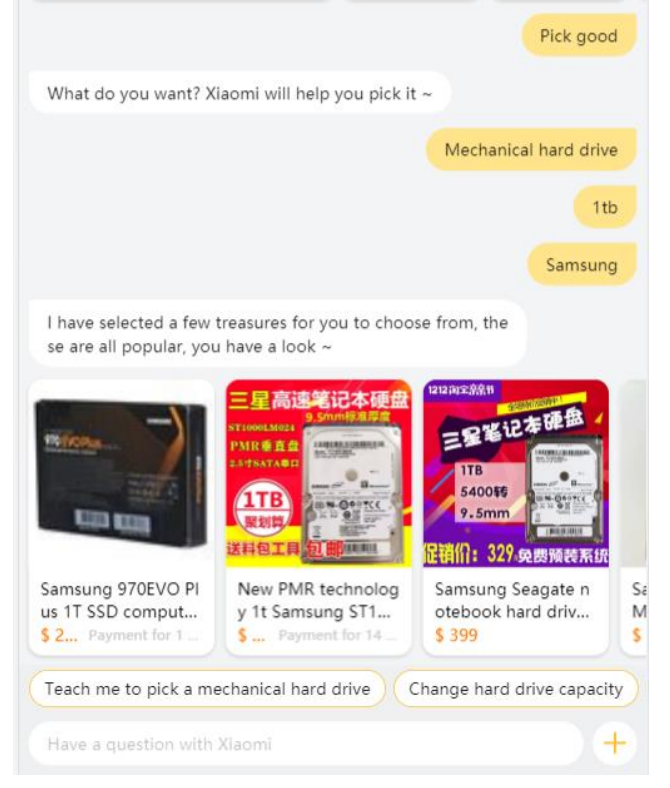}}
  \subfigure[Kuaishou]{
    \label{fig:subfig:b} 
    \centering
    \includegraphics[width=2in]{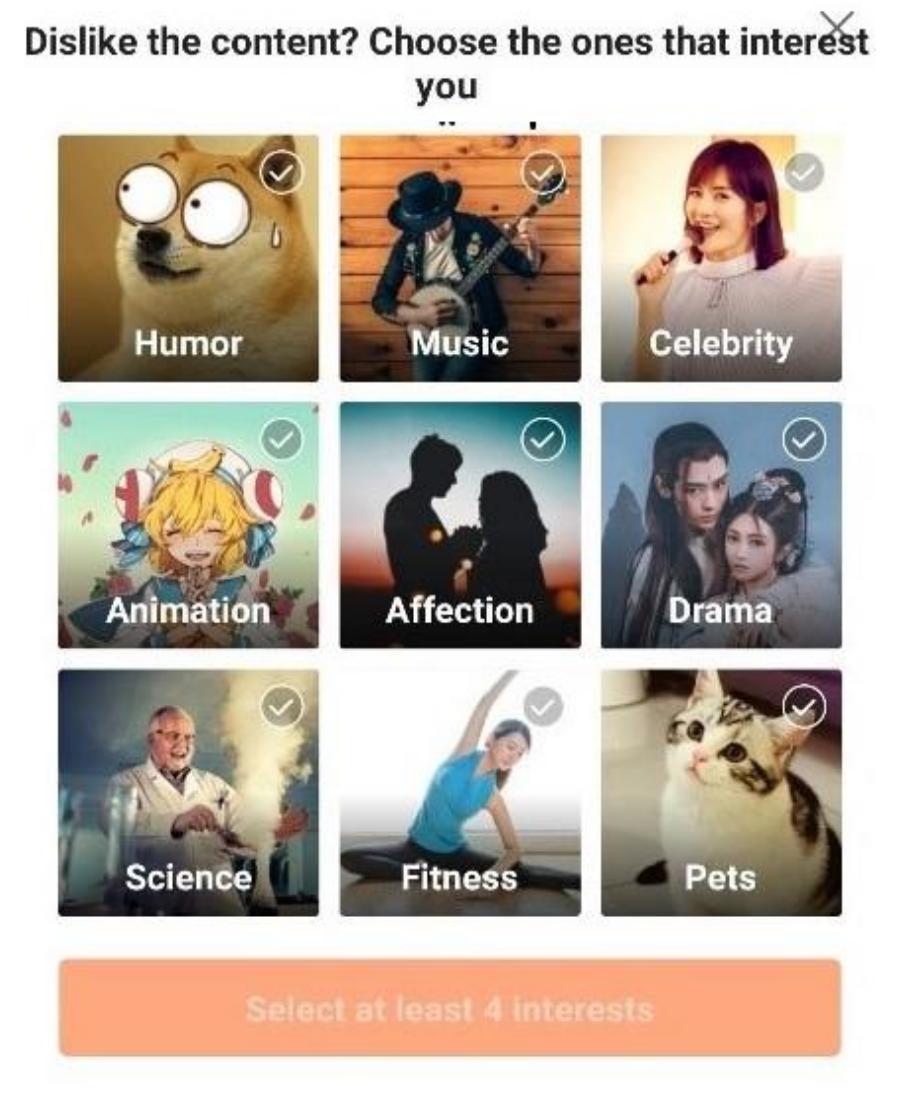}}
\caption{The chatbot interface in Taobao (a) and the pop-up box of attribute asking in Kuaishou (b).} 
\end{figure}

On the industry side, we have also witnessed an increasing trend on exploring the technology of CRS in various scenarios. 
A classical scenario is the customer service chatbot in E-commerce, which can help users find their interested products by dialogues. 
Figure~\ref{fig:subfig:a} shows the interface of the chatbot in Taobao\footnote{The largest Chinese E-commerce platform: https://consumerservice.taobao.com/}: when a user comes into the system, the bot can ask questions to elicit user preferences such as asking the user preferred attribute (e.g., \emph{brand} and \emph{categories}); then the bot makes recommendations according to the user's responses.

In addition to the chatbot, many scenarios that interact with users to facilitate information seeking can be abstracted as the problem of CRS. 
Taking Kuaishou app\footnote{A famous Chinese video-sharing platform: https://www.kuaishou.com/} as an example, when the user browses videos, the app will pop up a box to ask the user's favorite attributes of the videos (see Figure~\ref{fig:subfig:b}). This can be formalized as a conversational recommendation process~\cite{shimazu2002expertclerk, thompson2004personalized, lei20estimation, zhang2018towards, wei2015df}, which is characterized as form-based conversational recommendation system in~\cite{abs-2004-00646}. Popping up attribute box can be formalized as ''asking what attributes a user likes'' in the conversational recommendation while displaying videos can be formalized as making recommendation actions. 
Such mechanism has demonstrated to be effective hence been widely adopted in many applications like YouTube\footnote{A famous online video-sharing platform: https://www.youtube.com/}, Spotify\footnote{An international media services provider: https://www.spotify.com/us/}, Steam\footnote{The largest game platform: https://store.steampowered.com/}.

As an emerging topic, conversational recommendation has been explored under various settings with different focuses~\cite{christakopoulou2016towards, christakopoulou2018q,zhang2018towards,sun2018conversational, priyogi2019preference, yu2019visual, sardella2019approach, zhang2020toward,chen-etal-2019-towards}. 
Focusing on cold-start users, the first attempt is made by \cite{christakopoulou2016towards}, which identified the key of cold-start conversational recommendation as pursuing the exploration-exploitation (EE) trade-off. 
It addresses the problem with the multi-arm bandit approach~\cite{li2010contextual}, modeling items as the arms to achieve the EE trade-off. However, this method is inefficient by soliciting user preference only of the level of items, since the number of items can be million or even billion in real-world applications. 
To address this issue, a recent work \cite{zhang2020toward} integrates the attribute asking into the system. 
It models attributes and items as two different types of arms, using a manually-defined function to determine the timing of attribute asking (e.g., asking an attribute per three turns). We argue that a simple heuristic function is insufficient to find the best timing of attribute asking, which depends on many ad-hoc factors like the number of current turn and user's feedback. The strategy of attribute asking is critical for the usability and effectiveness of a CRS, which aims to identify the desired item(s) with the fewest turns, so as to avoid bothering the user with too many questions. However, it has not been seriously considered in conversational recommendation with EE trade-off.  

In this work, we explore the central theme of conversational recommendation for cold-start users. 
We highlight two desired properties for such a CRS: 
1) deciding the strategy of asking and recommending in an intelligent way, and
2) keeping the balance between exploiting known preferences and exploring new interests to achieve successful recommendations.
To this end, we propose the \textit{Conversational Thompson Sampling} (ConTS) method, modeling attribute asking and item recommending in the unified framework of Thompson Sampling.
The key idea is to model attributes and items as undifferentiated arms in the same arm space, where the reward estimation for each arm is jointly decided by the user's representation, candidate items, and candidate attributes to capture the mutual promotion of attributes and items. Through this way, we seamlessly unify attributes and items in conversational recommendation, leading to two main advantages of ConTS: (1) it addresses \emph{conversation policy questions} in CRS proposed in \cite{lei20estimation} --- what items to recommend, what attributes to ask, and whether to ask or recommend in a turn --- as the single problem of arm choosing, which is optimized for maximized rewards; and (2) it inherits the sampling and updating mechanism of contextual Thompson Sampling~\cite{agrawal2013thompson}, being able to achieve the EE balance naturally. 
The main contributions of this paper are as follows:
\begin{itemize}[leftmargin=*]
    \item We study the new task of conversational recommendation for cold-start users with attribute preference. By modeling attributes and items as indiscriminate arms in the same space, we propose a holistic solution named ConTS to solve the three conversation policy questions in CRS in an end-to-end manner. Meanwhile, we apply contextual Thompson Sampling to conversational recommendation to keep an EE balance in the cold-start scenario. 
    \item We conduct experiments on two existing datasets from Yelp and LastFM, and contribute a new dataset for cold-start conversational recommendation evaluation from Kuaishou (video-click records). 
    Extensive experiments show that our ConTS outperforms the state-of-the-art CRS methods \textit{Conversational UCB} (ConUCB)~\cite{zhang2020toward} and \textit{Estimation–Action–Reflection} (EAR)~\cite{lei20estimation} in both metrics of success rate and average turn for cold-start users. Further analysis shows the importance of exploration and the effectiveness of ConTS's action policy. All codes and data will be released to facilitate the research community studying CRS\footnote{https://github.com/xiwenchao/conTS-TOIS-2021}.
\end{itemize}

\section{Related works}\label{RW}
In the field of recommendation system, many static recommendation methods have achieved good performance. For example, matrix factorization~\cite{koren2009matrix}, factorization machines~\cite{FM}, NeuralFM~\cite{he2017neural} and DeepFM~\cite{guo2017deepfm} make use of historical user-item interaction records to estimate users' static preference based on collaborative filtering hypothesis. Recently some graph based recommendation models~\cite{vdberg2017graph, wang2019neural} attracts much attention for the ability of graph to describe complex relations between different agents. However, most of these works suffer from intrinsic problems for cold-start users as they need to be trained on a large offline dataset to make recommendation based on historical logs.

This limitation motivates research efforts in online recommendation in ~\cite{li2015online}. Among them, muti-arm bandit ~\cite{gentile2016context,wu2016contextual,wang2017factorization,wu2018learning,li2010contextual, auer2002finite, murphy2012machine} is a prominent type of approach. Bandit methods aim to maximize the accumulated reward for playing arms during a period of time by solving the exploit-explore problem. 
The most popular bandit algorithms can be divided into two categories: Thompson Sampling~\cite{graepel2010web, granmo2010solving, may2011simulation, murphy2012machine, agrawal2013thompson} and UCB~\cite{li2010contextual, auer2002finite, lai1985asymptotically}. Different from Thompson Sampling which samples from a changeable distribution, UCB keeps an upper confidence bound to keep EE balance. Some research works~\cite{russo2014learning, osband2017optimistic, chapelle2011empirical} have shown that Thompson Sampling has a better performance than UCB both theoretically and empirically. However, bandit algorithms only work in small arm pools, which usually requires a separate pre-processing on the candidate pool~\cite{li2010contextual, chapelle2011empirical}. It is because the large item pool makes it hard to efficiently hit the user-preferred item duration exploration.

Conversational recommendation systems (CRS)~\cite{christakopoulou2016towards, christakopoulou2018q,zhang2018towards,sun2018conversational, priyogi2019preference, yu2019visual, sardella2019approach, zhang2020toward,chen-etal-2019-towards} provides a promising solution. To enhance the performance of recommendation system with a large item pool, conversational recommendation ask questions to directly acquire the user's preference. Such information can help a CRS to adjust its recommendation strategy eﬀectively and enhance its performance. Extensive research eﬀorts have been recently devoted to explore this topic spreading various task formulations.
\citet{abs-2004-00646} classify CRS into two categories: NLP-based and form-based models. CRSs based on natural language~\cite{acl18/sequicity,jin2018explicit, jiang-2019-improving, ren-2020-thinking} are dynamic in terms of the possible dialogue ﬂow. Works in this direction focus on how to understand users’ preferences and intentions from their utterances in various forms and generate ﬂuent responses so as to deliver natural and eﬀective dialogues~\cite{acl18/sequicity,jin2018explicit, jiang-2019-improving, ren-2020-thinking}. 
In form-based CRS~\cite{averjanova2008map, hong2010interactive, dietz2019designing, mccarthy2004dynamic}, users typically follow pre-defined dialogue paths and interact with the application by filling forms through pre-defined options. These models are based on forms (e.g., buttons, radio buttons) and structured text for the output (e.g., in the form of a list of options). The example given in Section~\ref{intro}
of the Kuaishou app also belongs to this category. It requires the user to choose his favorite attributes listed in the box, which is an option pre-defined by the system. However, most of the models are designed for existing users instead of considering cold-start users with few historical data. A typical example is the EAR~\cite{lei20estimation}, which is trained on existing user's interaction records to estimate the preference of these users, thus fail to handle new users without historical information. 


There are pioneer works studying conversational recommendation problems for cold-start users. They basically leverage bandit methods to achieve EE balance. \citet{christakopoulou2016towards} systematically compare several bandit methods (i.e., Thompson Sampling, UCB, Greedy, Max Item Trait and Min Item Trait) and find that Thompson Sampling achieves the best performance. 
However, the resultant CRS is only able to recommend items to users but not able to ask question about the attributes. We argue that it is more efficient to explore the item space with the information of item attributes. In many real-world applications, the number of items could be very large-scale, e.g., there are over billions of products (videos) in Alibaba (YouTube). Soliciting user preference in the level of \textit{item attribute} is far more efficient than that at the level of \textit{item} ; thus if we know what attributes are preferred by the user now, we can safely filter out the items that do not contain the preferred attributes and facilitate efficient product searching. For example, if a user specifies that he wants to watch \textit{Science Fiction} movie, we can exclude other categories of movies for the consideration.


A very recent work by~\citet{zhang2020toward} further considers attribute asking in the framework of  UCB~\cite{li2010contextual, auer2002finite, lai1985asymptotically} for conversational recommendation, which helps to keep EE balance when recommending for cold-start users. It models the attributes and items separately as two set of arms in different arm spaces and decides whether to ask or recommend in a totally handcrafted way. Specifically, the authors define a frequency function $b(t)=5 * \left\lfloor {\log (t)} \right\rfloor$ where $t$ denotes the current conversation turn. ConUCB asks attributes only if $b(t)-b(t-1)>0$. However, we argue such handcrafted way is not robust.
In contrast, we model attributes and items as undifferentiated arms in the same arm space, and nicely fit the arm choosing in the framework of contextual Thompson Sampling. As such, all the three \emph{conversation policy questions} (e.g., what item to recommend; what attribute to ask; and whether to ask attribute or to recommend) are seamlessly modeled as the single problem of arm choosing. This results in a holistic model that is more robust and efficient than ConUCB.

\section{Preliminary}
This paper studies conversational recommendation for cold-start users. We choose the multi-round conversational recommendation (MCR) scenario~\cite{lei20estimation} as the problem setting because it is more reflective of real-world scenario~\cite{lei20estimation}. It is worth to mention that the MCR is a type form-based~\cite{abs-2004-00646} setting, which means users interact with the application by choosing from pre-defined options instead of dynamically generated natural languages. In this section, we first introduce the problem setting, and then sketch Thompson Sampling since our method is inspired from it. 
For the ease of understanding, the complex mathematics are skipped. We refer the reader to~\cite{graepel2010web, agrawal2013thompson} for more details. We give an intuitive illustration about how Thompson Sampling works in Section~\ref{supp} (Supplementary).
\begin{table}[t]
  \caption{A summary of main notations used in the paper}
  \label{tab:freq}
  \begin{tabular}{ccl}
    \toprule
    Notation  & Implication \\
    \midrule
    $u,\mathcal{U}$  & User and collection of all users\\
    $v,\mathcal{V}$  & Item and collection of all items\\
    $p,\mathcal{P}$  & Attribute and collection of all attributes\\
    $\mathbf{u}, \mathbf{v}, \mathbf{p}$ & Embedding of user, item and attribute\\
    $a, \mathbf{x}_a$  & Arm and embedding of arm $a$ in Bandit algorithms\\
    $r_a$ & Reward gotten from the user on arm $a$\\
    $\mathcal{A}$ & Collection of all arms\\
    $\textbf{B}_u, {\emph{\textbf{f}}_u}, {\bm{\mu} _u}$, $l$ & Parameters for user $u$ in contextual Thompson Sampling\\
    $\mathcal{P}_u$  & Currently known user $u$'s preferred attributes\\
    $\mathcal{P}_v$  & Item $v$' attributes\\
    $d$  & Dimension of embedding of user, item and attribute\\
    $k$  & The number of recommended items in one turn\\
    $T$  & The max turn of a conversation in a CRS\\
    $\mathcal{V}_k$ & Top $k$ items in the candidate item list\\
    $\mathbf{u}_{init}$ & Initialization of user embedding\\
    $\mathbf{u}_{orig}$ & User embedding in original contextual Thompson Sampling\\
    $r_a'$ & The de-biased reward in ConTS\\
    $I_{training}, I_{testing}, I_{validation}$ & The set for training, testing and validation\\
    
  \bottomrule
\end{tabular}
\end{table}


\subsection{Problem Setting}
In MCR, the system can perform attribute asking and item recommending multiple times until the recommendation is accepted or a maximum number of turns is reached. The objective is to achieve successful recommendations with the fewest conversation turns for each cold-start user $u$ who has no past historical interaction records. 
Let $\mathcal{V}$ denotes all the items and  $\mathcal{P}$ denotes all item attributes. 
Each item $v$ is described by a set of categorical attributes $\mathcal{P}_v$, e.g., the attributes of restaurants may include location, category, price level, and so on. 
In each turn $t$ ($t=1, 2, ..., T$,  where $T$ is the maximum number of turns) in a session, the system maintains two candidate pools, one for items $\mathcal{V}_t \subseteq \mathcal{V}$ and another for attributes $\mathcal{P}_t \subseteq \mathcal{P}$, from which we select attribute(s) to ask and top-$k$ items to recommend, respectively. Note that if it exceeds the maximum conversation turn $T$, we assume that the user will quit the conversation due to limited patience.
In the initial turn ($t=1$), the candidate pools contain all the items and attributes, which are gradually reduced with the proceeding of the conversation process. 

With $\mathcal{V}_t$ and $\mathcal{P}_t$ established in the beginning of turn $t$, the system takes an action to interact with the user --- either asking for attribute(s) or recommending $k$ items. Then the user will give feedback to the system, which can either be 1) like the attribute(s),  2) dislike the attribute(s)\footnote{More comprehensive than \cite{lei20estimation}, here we explicitly model user negative feedback on attributes.}, 3) accept the recommendations, or 4) reject the recommendations. Only if the user accepts the recommendations, the session ends successfully; In other cases, the session continues and the system needs to adjust the candidate pools, and possibly, the action policy. In terms of the policy for making actions, there are three questions to consider: 
(1) what attributes to ask, (2) what items to recommend, and (3) whether to ask or recommend in a turn. 
Instead of addressing each question with a separate component~\cite{lei20estimation,sun2018conversational}, our method seamlessly addresses them in a holistic framework. 

\subsection{Thompson Sampling}\label{TS1}
As our method is inspired by Thompson Sampling, we here briefly introduce it. 
Thompson Sampling~\cite{graepel2010web, granmo2010solving, may2011simulation, murphy2012machine, agrawal2013thompson} is a typical class of bandit algorithms ~\cite{russo2018tutorial} which balance the exploration and exploitation during online learning. 
Here, the exploitation means taking actions according to the model's current estimation of the user's interest, and exploration means exploring a user's unknown preference to pursue a possibly higher reward in the future. 
Differently from bandit algorithms that use the upper confidence bound (UCB)~\cite{UCB1,li2010contextual} to model the uncertainty of the learner's estimation on a user's preference, Thompson Sampling balances EE by sampling from a posterior distribution of the reward estimation, which is shown to be a more effective and robust method in many situations according to some previous research works~\cite{chapelle2011empirical,russo2014learning}.

In the typical setting of contextual bandit problem, there is an arm pool $\mathcal{A}$ and a set of contexts $\mathcal{X}$. In an interaction turn, the model needs to select an arm ${a \subseteq {\mathcal{A}}}$ to play given a context ${\bx \subseteq {\mathcal{X}}}$, and then gets a reward $r_a$ from the users accordingly. The goal of the arm choosing is to maximize the accumulated reward given the limited interaction turns. In the interactive recommendation scenario, an arm $a$ can represent an item and the context $\bx$ can represent the accessible information for the recommendation decision-making,
such as the arm (i.e., item) embeddings. 
In a turn, the recommender chooses an item to display to the user (i.e., play an arm) and gets user feedback as the reward, either by accepting (positive reward) or rejecting (negative reward). In the linear contextual bandit setting, the reward $r_{a}$ for an arm $a$ under context $\mathbf{x}_{a}$ is assumed to be generated from
an (unknown) distribution with mean $\mathbf{u}^{\mt}\mathbf{x}_{a}$, where $\mathbf{u} \in \mathbf{R}^d$ is a fixed but unknown parameter. In the recommendation scenario, this unknown variable $\mathbf{u}$ can be considered as an embedding that reflects a users' preference. 

In Thompson sampling, with a prior distribution $P(\mathbf{u})$ on the unknown parameter $\mathbf{u}$ and a collection of observed triples $\mathcal{D}_t = \{(\bx_1;a_1;r_{a(1)}),$ $ (\bx_2;a_2;r_{a(2)}), ... (\bx_{t-1};a_{t-1};r_{a(t-1)})\ \}$ in the current session at time $t$, a posterior estimation $P(\bu)$ of the user's preference can be derived  according to the Bayes rule:
\begin{equation}
\begin{aligned}
  P(\bu|\mathcal{D}_t) \propto P(\mathcal{D}_t|\bu)P(\bu) ,
\end{aligned} 
\end{equation}
where $P(\mathcal{D}_t|\bu)$ is the likelihood function.

Thompson sampling samples $\tilde \bu$ from $P(\bu|\mathcal{D}_t)$ to get instantiated parameters in the turn, based on which, the reward for each arm $a$ is calculated as $\tilde r_{a} = \tilde \bu^{\mt} \bx_{a}$. An arm is chosen to play according to the reward calculated from the sampled parameter $\argmax_{a} \tilde \bu^{\mt} \bx_{a} $. After getting the user's feedback, we have a new triple added into $\mathcal{D}_t$. In the process, sampling and updating are the key mechanisms to ensure the EE balance. We next use an example to illustrate it.

A commonly used likelihood function $P(\mathcal{D}_t|\bu)$ and prior $P(\bu)$ are Gaussian likelihood function and Gaussian prior~\cite{agrawal2013thompson}, respectively.  Contextual Thompson Sampling~\cite{agrawal2013thompson} naturally assumes the embedding for user $u$ also follows a Gaussian distribution, which can be written as $\mathcal{N}(\bm{\mu}_u,{l^2}\textbf{B}_u^{ - 1})$. The mean ${\bm{\mu} _u}$ denotes the estimated expectation of user embedding \textbf{u} and the covariance matrix ${l^2}\textbf{B}_u^{ - 1}$ denotes the uncertainty about the embedding (note that $\textbf{B}_u$ is the inverse covariance matrix), and $l$ is a hyper-parameter to control the uncertainty.
After the user gives the feedback (reward) $r_a$ on arm $a$, the model updates the parameters as:
\begin{equation}
\begin{aligned}
   \textbf{B}_u =  \textbf{B}_u + \textbf{x}_{a(t)}{\textbf{x}_{a(t)}}^T
\end{aligned} 
\end{equation}
\begin{equation}
\begin{aligned}
   {\emph{\textbf{f}}_u} = {\emph{\textbf{f}}_u} + r_a * {\textbf{x}_{a(t)}}
\end{aligned} 
\end{equation}
\begin{equation}
\begin{aligned}
   {\bm{\mu} _u} = \textbf{B}_u^{ - 1}{\emph{\textbf{f}}_u} ,
\end{aligned} 
\end{equation}
where $\textbf{x}_{a(t)}$ is the embedding of the chosen arm $a(t)$ at turn $t$, $\textbf{B}_u$ is initialized as the identity matrix and ${\emph{\textbf{f}}_u}$ as a zero vector. Then contextual Thompson Sampling will sample from the posterior distribution $\mathcal{N}(\bm{\mu}_u,{l^2}\textbf{B}_u^{ - 1})$ to get an estimation of user embedding $\textbf{u}$ under uncertainty.

It will keep updating $\mathcal{N}(\bm{\mu}_u,{l^2}\textbf{B}_u^{ - 1})$ during the interaction process. On the one hand, this posterior update incorporates past experience into the reward estimation; on the other hand, it reflects the model's uncertainty on arms with different context.
For example, if the user gives a negative feedback to an arm $a$, contextual Thompson Sampling tends to score lower for $a$ and other arms of the similar embedding with $a$ by updating the mean $\bm{\mu}_u$. And whenever the user gives feedback to an arm, contextual Thompson Sampling will update $\textbf{B}_u$ to discourage further exploration of this arm because we have already known the user's preference on it. More details about the strategy of update is given in Section~\ref{supp}.


\section{method}\label{method}

As discussed in Section~\ref{intro}, the key challenges in conversational recommendation for cold-start scenario lie on two aspects: (1) the strategy to decide whether to ask or recommend; and (2) the balance of EE. To tackle the two challenges, we propose to unify the attributes and items seamlessly into the one arm space, resulting in our ConTS model. Specifically, it comprises three parts, namely 1) Initialization and Sampling, 2) Arm Choosing, and 3) Updating. We will detail them after a model overview in this section.

\subsection{Overview}

\begin{figure}
\begin{minipage}[t]{0.8\linewidth}
    \includegraphics[width=\linewidth]{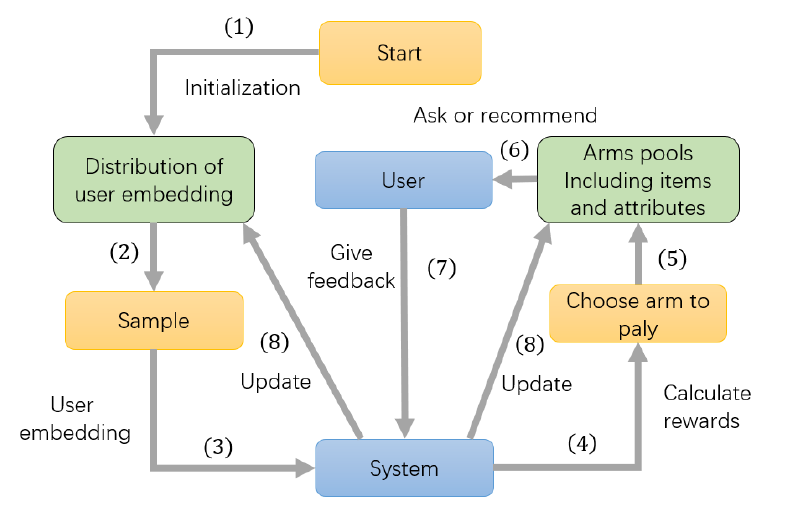}
\end{minipage}%

\caption{The example of ConTS' actions during a whole conversational recommendation process for cold-start users. Orange blocks represent operations, green blocks represent parameters and blue blocks resemble agents in the scenario.} 
\label{f2}
\end{figure}

Figure~\ref{f2} shows the workflow of our ConTS in a conversation session. We overview the model by going through the major components (the numbers in text description correspond to the numbers in Figure~\ref{f2}).

\noindent \textbf{Initialization and Sampling}: If it is the first turn in the session, we need to initialize the distribution of user embedding by averaging the pre-trained embedding of existing users (step (1)\textasciitilde(2)). Then, the conversation cycle starts from sampling an instantiated embedding for the user from the currently estimated distribution (step (3)). The sampling step exactly follows the contextual Thompson Sampling~\cite{agrawal2013thompson}, { leveraging its advantage to achieve EE balance. }
 

 
\noindent \textbf{Arm Choosing}: ConTS decides an action (asking or recommending) by choosing an arm with the goal to maximize the reward (step (4)\textasciitilde(5)). Here, we make a key contribution to unify the attributes ${p \subseteq {\mathcal{P}}}$ and items ${v \subseteq {\mathcal{V}}}$ as the undifferentiated arms in the same space (the whole arm pool is ${{\mathcal{A}} = {\mathcal{P}} \cup {\mathcal{V}}}$). Our ConTS only needs to choose arms based on a unified reward estimation function which captures the mutual promotion of attributes and items. As such, the \emph{conversation policy questions} -- what items to recommend, what attributes to ask, and whether to ask or recommend in a turn -- are reduced to a single question of arm choosing in ConTS. If the chosen arm is an item, a CRS recommends the top $k$ items; and if the arm is an attribute, a CRS asks the attribute (or several top attributes, depending on the problem setting, \emph{c.f.} Section~\ref{DS}) (step (6)).

We argue this is the key difference between our ConTS and ConUCB~\cite{zhang2020toward}. In ConUCB, the items and attributes are separately modelled in two different arm spaces. The problems of \emph{what item to recommend} and \emph{what attribute to ask} are modeled independently where each requires specific efforts and the question of \emph{whether to ask or recommend in a turn} is addressed by an ad-hoc rule (e.g., asking attributes each five turns).




\noindent \textbf{Updating}: After playing an arm (recommend item or ask attribute), the system will receive feedback from the user (step (7)). 
It firstly updates the current arm pool according to the feedback (e.g., removing items not containing the user's preferred attributes, see Section~\ref{US}). ConTS then updates the distribution of user embedding based on our unified reward estimation function and the user's feedback (step (8)). We will detail it in Section~\ref{US}. 



\subsection{Initialization and Sampling}\label{IS}

\hspace*{0.4cm}\textbf{Initialization with offline FM}. As the standard step in contextual Thompson Sampling, we need to get the arm embedding. We train an offline Factorization Machine (FM) model  following EAR~\cite{lei20estimation} on records of existing users to simultaneously get the embeddings of all attributes, items and existing users. Specifically, we train the FM by Bayesian Personalized Ranking (BPR)~\cite{BPR}, aiming to make it rank user's preferred items and attributes higher than the others. We do muti-task training, jointly training on the two task of item prediction and attribute prediction. Specifically, we first train the model for item prediction. After it converges, we continue to optimize the FM for attribute prediction. The embeddings of all attributes and items are in the same embedding space, which will be treated as arm embeddings later in Section~\ref{RA}. 

\textbf{Online posterior sampling}. ConTS follows the contextual Thompson Sampling described in Section \ref{TS1}, by assuming the embedding for a new user $u$ follows a multidimensional Gaussian distribution $\mathcal{N}(\bm{\mu}_u,{l^2}\mathbf{B}_u^{ - 1})$. Conventionally, ${\mathbf{B}_u}$ and $\bm{\mu}_u$ are initialized with identity matrix and zero vector separately. However, we hypothesize that the historical action of existing users can help to estimate the preference of new users. Without any previous information, it is reasonable to assume that the preference of new users to be the average of exsing users'. Thus, { we initialize $\bm{\mu}_u$ as the average embedding of existing users} while following the convention to initialize ${\mathbf{B}_u}$ as identity matrix. Specifically, if $\mathcal{U}^{old}$ denotes the collection of all embeddings of existing users, then,
\begin{equation}
\label{init}
\begin{aligned}
   \mathbf{u}_{init} =  \frac{{1}}{N}\sum_{i=1}^N \mathbf{u}_i , \mathbf{u}_i \in \mathcal{U}^{old}.
\end{aligned} 
\end{equation}

Correspondingly, the intermediate variable ${\emph{\textbf{f}}_u}$ is also initialized  with $\mathbf{u}_{init}$ since ${\emph{\textbf{f}}_u}$ is updated by Eq~\ref{update4} (i.e., ${\bm{\mu} _u} = \mathbf{B}_u^{ - 1}{\emph{\textbf{f}}_u}$, where $\mathbf{B}_u$ is initialized by identity matrix).

After initialization, ConTS starts a MCR session to interact with the user.
At the beginning of each turn $t = 1,2,...,T$ in a MCR session, ConTS samples from $\mathcal{N}(\bm{\mu}_u,{l^2}\mathbf{B}_u^{ - 1})$ to get user embedding $\tilde \bu$\footnote{Note that both $\bm{\mu}_u$ and ${l^2}\mathbf{B}_u^{ - 1}$ are changed in each turn according to user feedback (see Section~\ref{US}).}. 
The sampling is the key step to { achieve EE} balance. On one hand, ConTS uses the mean $\bm{\mu}_u$ to control the expectation of the sampling result to exploit user's currently known preference. On the other hand, ConTS uses covariance ${l^2}\mathbf{B}_u^{ - 1}$ to model the uncertainty about the estimated user preference, which decides how ConTS explore user's latent unknown preference.

\renewcommand{\algorithmicrequire}{\textbf{Input:}}
\renewcommand{\algorithmicensure}{\textbf{Output:}}
\algnewcommand{\algorithmicand}{\textbf{ and }}
\algnewcommand{\algorithmicor}{\textbf{ or }}
\algnewcommand{\OR}{\algorithmicor}
\algnewcommand{\AND}{\algorithmicand}
\begin{algorithm}[tb]  
	\caption{\textbf{Conversational Thompson Sampling (ConTS)}}			
	\label{alg:alg1}	
	\begin{algorithmic}[1]	
	\Require			
		user $u$,
        the set of all attributes $\mathcal{P}$,
        the set of all items $\mathcal{V}$,
        the number of items to recommend $k$,
        maximum turn number $T$,
        hyper parameter $l$;
	\Ensure Recommendation result: success or fail;
    \State Initialization: for a new user $u$, do the following initialization:\\
    $\mathcal{A}_{u}=\mathcal{P}\cup\mathcal{V};~$
    $\mathbf{u}_{init} =  \frac{{1}}{N}\sum_{i=1}^N [ \mathbf{u}_i | u_i \in \mathcal{U}^{old}];~$
    ${\bm{\mu} _u} = {\emph{\textbf{f}}_u} = \textbf{u}_{init};~$ ${\textbf{B}_u} = {\textbf{I}_d};~$ $\mathcal{P}_{u}=$$\emptyset;$
    
    \For{Conversation turn $t=1,2,3...T$}
        \State Sample $\tilde \bu$ from the distribution:
        $\tilde{\textbf{u}} \sim \mathcal{N}(\bm{\mu}_u,{l^2}\textbf{B}_u^{ - 1})$

        \State Play arm ${a(t) = argmax_{a \subset {\mathcal{A}_u}}~ \tilde{\textbf{u}^{\mt}} \mathbf{x}_{a}
    + \sum_{p_i\in\mathcal{P}_u} {\mathbf{x}_a}^T \mathbf{p}_i}$
        
        \If{$a(t) \in \mathcal{P}$}
            \State ask $u$'s preference on $a(t)$
            \State get $u$'s feedback $r_a$
            \If{$u$ accepts $a(t)$}
                \State $\mathcal{P}_{u}=\mathcal{P}_{u} \cup a(t)$ 
                \State $\mathcal{A}_{u}=\{a\left|{a\in (\mathcal{P} \setminus \mathcal{P}_{u})~or~(a\in \mathcal V} \right.\& ~a(t)\in \mathcal{P}_{u})\}$
            
                \Else: $\mathcal{A}_{u} = \mathcal{A}_{u} \setminus a(t)$
            \EndIf
        \Else{$~a(t) \subset \mathcal{V}$}
            \State recommend top k items $\mathcal{V}_{k}$ from ranked item pool $\mathcal{V}_{rank}$
            \State get $u$'s feedback $r$
            \If{$u$ accepts $\mathcal{V}_{k}$}
                \State \emph{Recommendation succeeds, Quit}; 
            \Else~User rejects $\mathcal{V}_{k}$
                \State Update: $\mathcal{A}_{u} = \mathcal{A}_{u} \setminus \mathcal{V}_{k}$
            \EndIf
        \EndIf
        \State Update Bandit parameters in sequence by:
                \State $\qquad \textbf{B}_u =  \textbf{B}_u + \textbf{x}_{a(t)}{\textbf{x}_{a(t)}}^T$
                \State $\qquad r_a' = r_a - {\textbf{x}_{a(t)}}^T (\textbf{u}_{init} + \sum_{p_i\in\mathcal{P}_u} \mathbf{p}_i)$
                \State $\qquad {\emph{\textbf{f}}_u} = {\emph{\textbf{f}}_u} + r_a' * {\textbf{x}_{a(t)}}$
                \State $\qquad {\bm{\mu} _u} = \textbf{B}_u^{ - 1}{\emph{\textbf{f}}_u}$
        \If{$t=T$}
            \State \emph{Recommendation fails, Quit};
        \EndIf
    
    \EndFor
	\end{algorithmic}
	\label{algo:MrCRS}
\end{algorithm}
\vspace{2pt}

\subsection{Arm Choosing}\label{RA}

Once obtaining user embedding by posterior sampling, the agent needs to take an action, either by asking an attribute or recommending items. This is much more complex than existing contextual Thompson Sampling methods where the action is simply to choose an item to recommend. In the MCR scenario, a CRS needs to consider more questions: 1) what attributes to ask, 2) what items to recommend, and 3) whether to ask or recommend in a turn. To address those problems, ConTS adopts a simple but efficient strategy to model all the items and attributes as  undifferentiated arms in the same arm space. As such, the aforementioned three problems in the MCR scenario reduced to the problem of arm choosing, which can be nicely solved by the \emph{maximum rewards} approach in contextual Thompson Sampling. Specially, we assume the expected reward of arm $a$ of user $u$ is generated by:

\begin{equation}
\label{rewards}
\begin{aligned}
  r(a,u,\cP_u) \sim \cN(  \bu^{\mt} \bx_{a}
    + \sum_{p_i\in \cP_u} {\mathbf{x}_a}^T \bp_i, l^2) ,
\end{aligned} 
\end{equation}

\noindent where $l$ is the standard derivation of the Gaussian noise in the reward observation. $\bu$ and $\textbf{x}_{a}$ represent the embedding of the user and arm, and $\mathcal{P}_{u}$ denotes user's currently known preferred attributes obtained from the conversation by asking attributes.  It explicitly takes the attributes confirmed by the user during a conversation as a strong indicator for user preference. The first part $\bu^{\mt}\mathbf{x}_{a}$ models the general preference of user $u$ to arm $a$, and the second part $\sum_{p_i\in\mathcal{P}_u} {\mathbf{x}_a}^T \mathbf{p}_i$ models the affinity between arm $a$ and the user's preferred attributes $\mathcal{P}_{u}$. This formula is different from the linear reward generation assumption in contextual Thompson Sampling in the sense that it explicitly capture the mutual promotion of items and attributes. We intentionally design this formula to exactly follow EAR~\cite{lei20estimation}. This is because we want to establish a fair comparison between our ConTS and EAR to demonstrate the effectiveness of EE balance of the Thompson Sampling framework in the conversational recommendation for cold-start users (See Section~\ref{EX}).




Based on Eq.~\ref{rewards}, the system will choose the arm with the maximal reward to play. That is to say, if the chosen arm represents an attribute, the model will ask the user's preference about the attribute (or several top attributes). However, if the chosen arm is an item, the model will directly recommend the top $k$ items with max rewards to the user. 
{ The arm-choosing policy seamlessly unifies the asking attributes and recommending items together in an holistic framework, while ConUCB~\cite{zhang2020toward} separately models attributes and items, and decides whether to ask or recommend based on a hand-crafted heuristics.}
\vspace{2pt}


\subsection{Updating}\label{US}

After asking a question or recommending items, the agent will receive feedback from the user and the feedback will be used to update our model.
Firstly, ConTS renews the arm pool in the current conversation. If the user gives a negative feedback to either attribute(s) or items, we remove them from the current arm pool $\mathcal{A}_{u}$. If he likes the attribute, it is added into the user's preference attribute pool $\mathcal{P}_{u}$ and the agent {filters out items} without the preferred attribute from candidate pool. And if the user accepts the recommended item, the session successfully ended according to the MCR scenario. 

Different from the typical update methods in contextual Thompson Sampling described in Section~\ref{TS1} in Eq. (\ref{update2}), ConTS is adapted to conversational recommendation scenario for cold-start users by considering two additional factors: $\mathbf{u}_{init}$ for user embedding initialization and $\sum_{p_i\in\mathcal{P}_u}$ for reward estimation. In fact, Eq. (\ref{rewards}) can be re-written as:

\begin{equation}\label{eq12}
\begin{aligned}
r(a,u,\mathcal{P}_u) \sim\ &  \cN(  \bu^{\mt}\mathbf{x}_{a(t)} + \sum_{p_i\in\mathcal{P}_u} {\mathbf{x}_{a(t)}}^T \mathbf{p}_i, l^2) = \cN \big({\textbf{x}_{a(t)}}^T (\textbf{u}_{orig} + \textbf{u}_{init} + \sum_{p_i\in\mathcal{P}_u} \mathbf{p}_i) , l^2 \big),
\end{aligned}
\end{equation}

\noindent where $\mathbf{u}_{orig}$ denotes the user embedding exactly in the original contextual Thompson Sampling~\cite{agrawal2013thompson}, in which $\mathbf{u}_{orig}$ is initialized with zero vector, and $r(a,u,\mathcal{P}_u)$ is our observed reward according to user's feedback. During the whole process, $\mathbf{u}_{orig}$ is the only variable to be updated while $\mathbf{u}_{init}$ and $\sum_{p_i\in\mathcal{P}_u} \mathbf{p}_i$ are given, which can be regarded as the bias used in predicting the real reward from the original user embedding. According to Eq. (\ref{eq12}), we have ${\textbf{x}_{a(t)}}^T \textbf{u}_{orig} = 
   \bbE[r(a,u,\mathcal{P}_u)] - {\textbf{x}_{a(t)}}^T (\textbf{u}_{init} + \sum_{p_i\in\mathcal{P}_u} \mathbf{p}_i)$. Intuitively speaking, the reward generated by the user embedding $u_{orig}$ needs to be de-biased from the observed reward $r_a$ by removing the effect of the two bias terms. This intuition can also be reflected in the derivation of the user embedding's posterior distribution as shown in Eq.~\eqref{gauss_post}. 
   To simplify the notations, we define $r_{a}' =  r_a - {\textbf{x}_{a(t)}}^T (\textbf{u}_{init} + \sum_{p_i\in\mathcal{P}_u} \mathbf{p}_i)$, $\Sigma = {l^2} \bB_u^{ - 1}$, and simplify $\bu_{orig}$ as $\bu_{o}$.  
\begin{align} \label{gauss_post}
 P(\bu_o|(\bx, & a, r_{a})) \propto   P((\bx, a, r_{a})| \bu_o) P(\bu_o) \\ \nonumber
  & =  \cN(\bu_{o}^{\mt} \bx_{a} + \bu_{init}^{\mt} \bx_{a}
    + \sum_{p_i\in \cP_u} {\mathbf{x}_a}^T \bp_i, l^2) \cN(\bm{\mu}_u, \Sigma) \\  \nonumber
    & = \frac{\text{det}(\Sigma)^{-1/2}}{\sqrt{(2 \pi )^{d+1} \sigma^2}}   \exp(-\frac{(r_{a} - \bu_{o}^{\mt} \bx_{a} - \bu_{init}^{\mt} \bx_{a} 
    - \sum_{p_i\in \cP_u} {\mathbf{x}_a}^T \bp_i)^2}{2l^2}) \exp(-\frac{ (\bu_o - \bm{\mu}_u)^\mt \Sigma^{-1} (\bu_o - \bm{\mu}_u) }{2}) \\ \nonumber
    & =  \frac{\text{det}(\Sigma)^{-1/2}}{\sqrt{(2 \pi )^{d+1} \sigma^2}}   \exp(-\frac{(r_a' - \bu_{o}^{\mt} \bx_{a})^2 }{2l^2}) \exp(-\frac{ (\bu_o - \bm{\mu}_u)^\mt \Sigma^{-1} (\bu_o - \bm{\mu}_u) }{2}) \\ \nonumber 
    & = \frac{\text{det}(\Sigma)^{-1/2}}{\sqrt{(2 \pi )^{d+1} \sigma^2}}    \exp(\frac{\bu_o^\mt ( \Sigma^{-1} + \frac{1}{l^2}  \bx_{a} \bx_{a}^\mt ) - 2\bu_o^\mt ( \Sigma^{-1} \bm{\mu}_u + \frac{1}{l^2}  \bx_a r_a') +  \bm{\mu}_u^{\mt} \Sigma^{-1} \bm{\mu}_u + \frac{{r_a'}^2}{l^2}   }{-2}) \\ \nonumber
\end{align}

Since Gaussian distribution has conjugate prior, the posterior $P(\bu_o|(\bx, a, r_{a}))$ still follows the Gaussian distribution, i.e., it can be written as $P(\bu_o|(\bx, a, r_{a})) = \cN(\mu_{o, \text{post}}, l^2 \Sigma_{o, \text{post}} )$, in which $ \bu_{o, \text{post}}$ and $l^2 \Sigma_{o, \text{post}}$ are the posterior mean and covariance matrix respectively. According to Eq~\eqref{gauss_post}, it is easy to obtain,
\begin{align}\label{eq:post_cov}
   & \Sigma^{-1}_{o, \text{post}} =   \Sigma^{-1} + \frac{1}{l^2} \bx_{a} \bx_{a}^\mt = \frac{1}{l^2} ( \bB_{u} +  \bx_{a} \bx_{a}^\mt) \\ 
    & \bu_{o, \text{post}} = \Sigma_{o, \text{post}} ( \Sigma^{-1} \bm{\mu} + \frac{1}{l^2} \bx_a r_a') = \frac{1}{l^2}  \Sigma_{o, \text{post}} ( \bB_u \bm{\mu}_u + \bx_a r_a') = \frac{1}{l^2}  \Sigma_{o, \text{post}} (\bf_{u} + \bx_a r_a')  \label{eq:post_mean}
\end{align}

According to the derived posterior mean and covariance matrix in Eq.~\eqref{eq:post_cov} and Eq.~\eqref{eq:post_mean}, ConTS updates parameters of the distribution for the user's embedding as follows:
\begin{equation}
\label{update1}
\begin{aligned}
   \textbf{B}_u =  \textbf{B}_u + \textbf{x}_{a(t)}{\textbf{x}_{a(t)}}^T
\end{aligned} 
\end{equation}
\begin{equation}
\label{update2}
\begin{aligned}
   r_a' = r_a - {\textbf{x}_{a(t)}}^T (\textbf{u}_{init} + \sum_{p_i\in\mathcal{P}_u} \mathbf{p}_i)
\end{aligned} 
\end{equation}
\begin{equation}
\label{update3}
\begin{aligned}
   {\emph{\textbf{f}}_u} = {\emph{\textbf{f}}_u} + r_a' * {\textbf{x}_{a(t)}}
\end{aligned} 
\end{equation}
\begin{equation}
\label{update4}
\begin{aligned}
   {\bm{\mu} _u} = \textbf{B}_u^{ - 1}{\emph{\textbf{f}}_u} .
\end{aligned} 
\end{equation}


\subsection{Complexity Analysis}
Lastly, we briefly analyse the complexity of ConTS. If $n = |{\mathcal{V}_t}|$ denotes the length of current candidate item pool, $k = |{\mathcal{P}_t}|$ denotes the length of the current candidate attribute pool, and $d$ denotes the length of arm's embedding, the complexity of ConTS is $O(2 * d * (n + k) + d^2)$. The first part $O(2 * d * (n + k)$ is the complexity of ranking all items and attributes (usually $n >> k$), which is general in most of traditional statistic recommendation models such as matrix factorization~\cite{koren2009matrix} and factorization machines~\cite{FM}.
The second part $O(d^2)$ is the complexity of calculating the inverse of matrix of user embedding, which is a typical part of complexity for contextual Thompson Sampling algorithms~\cite{li2010contextual, agrawal2013thompson}. Note that the pre-trail computational complexity of matrix inverse can be reduced to $O(d^2)$ according to~\cite{li2010contextual}.
Since ConTS is based on contextual Thompson Sampling, the detailed analysis of complexity of ConTS is the same as contextual contextual Thompson Sampling. Therefore, we directly give the result as readers can easily find the detailed analysis on the classic literature about contextual Thompson Sampling~\cite{li2010contextual, agrawal2013thompson}. To sum up, our ConTS does not involve any higher-order complexities compared to standard static recommendation methods as well as Contextual Thompson Sampling methods, being practicable in real applications.

\section{experiments}\label{EX}


In this section, we conduct experiments to compare ConTS with representative conversational recommendation systems and the three variants of ConTS, each with one key component dropped in the original design. We do not only examine the overall performances, but also analyze the key mechanisms designed in ConTS e.g., keeping EE balance, seamlessly unifying items and attributes, etc.
Moreover, we design experiments to discuss ConTS's advantages over ConUCB and compare the performance of Thompson Sampling and UCB in our setting. Furthermore, we explore the performance of all methods on smaller max conversation turns. Lastly, we conduct experiments to study the impact of different cold-start degrees and do a case study for comparison.

\subsection{Datasets}\label{DS}
We perform experiments on three real-world datasets which vary significantly in domains and settings. The settings include enumerated questions on Yelp, binary questions on LastFM and multi-attribute questions on Kuaishou (details are shown in the following). We compare ConTS with other methods on the three different settings, aiming to evaluate the model's generalization in various conversational recommendation scenarios. 

\begin{itemize}
    \item \textbf{Yelp}\footnote{https://drive.google.com/open?id=13WcWe9JthbiSjcGeCWSTB6Ir7O6riiYy \label{web}}: This is a dateset for business recommendation. We use the version processed by~\cite{lei20estimation}, which contains a two-level taxonomy on the attributes in the original Yelp data\footnote{https://www.yelp.com/dataset/}. For example, the original attributes {"wine", "beer", "whiskey"} all belong to the parent attribute "alcohol". There are 29 such parent attributes over the 590 original attributes. The agent will choose from these 29 attributes to ask during a conversation. Meanwhile, all the child attributes will be displayed to the user for himself to choose the preferred one. Similar to EAR~\cite{lei20estimation}, this setting is denoted as \textit{enumerated question}. This step is to avoid overly lengthy conversation caused by a large attribute space.
    
    \item \textbf{LastFM}\footref{web}: This dataset is for music artist recommendation. We also follow exactly the pre-processing by~\cite{lei20estimation}. Different from the setting in Yelp dataset, this dataset is designed for binary questions. This means the that the system chooses one of the 33 attributes and ask the user to indicate the binary feedback (like or dislike) on the asked attribute.
    
    \item \textbf{Kuaishou}\footnote{https://github.com/xiwenchao/Kuaishou-data/}:  We construct a new dataset built on the video-click records of cold-start users in Kuaishou app. As shown in Figure~\ref{fig:subfig:b}, the app will show an interface to the user and ask for his preference on multiple attributes instead of a single one. Following this scenario, each time when the agent decides to ask for preference on attributes, our ConTS chooses the top 12 attributes according the estimated rewards. We still ask the user to give binary feedback to each attribute. It is also worth mentioning that we follow~\cite{lei20estimation} to filter out low-frequency data, including the users with less than 10 view records and the attributes with less than 5 occurrences. This is to make sure we can simulate more than 10 conversations for each user such that our evaluation is more robust.
\end{itemize}

\subsubsection{Dataset Partition}
The statistics of the three datasets are summarized in Table~\ref{t0}. For each dataset, we partition it into a training set, a testing set and a validation set. As discussed in Section~\ref{RA}, the training set is used to train the embedding of existing users, items and attributes offline. And the testing set is utilized for the conversation simulation of cold-start users which is detailed in Section~\ref{user_simu}. The validation set is used for parameter tuning.

To simulate the cold-start setting, the training set, testing sets and validation set are user-disjoint. This means that, for any user in one set, we cannot find any corresponding user-item interaction records in the other two sets. As such, we treat the users in the training set as existing users while the users in validation and testing set as new users as their interaction records do not appear in the training set. Note that the partition is based on the proportion of interaction records. Specifically, we use Algorithm~\ref{algo:partition} to generate such partition so that the user-item interaction records in training set take up the \textbf{70\%} of the all records in the whole dataset while the testing and validation set each take up \textbf{15\%} respectively. We choose this partition strategy to make our offline training of FM model to be on par with \cite{lei20estimation} where the FM model is trained using \textbf{70\%} of the user-item interaction records. The statistics of the training set, testing set and validation set of the three datasets are given in Table~\ref{t_new}.

\renewcommand{\algorithmicrequire}{\textbf{Input:}}
\renewcommand{\algorithmicensure}{\textbf{Output:}}
\begin{algorithm}[tb]  
	\caption{\textbf{Partition of Datasets}}			
	\label{alg:alg1}	
	\begin{algorithmic}[1]	
	\Require			
		collection of all users $U$,
		collection of all user-item interaction records $I$,
		collection of all interaction records for any user $I_u$ where $u \in U$;
	\Ensure training set $I_{training}$,
        testing set $I_{testing}$, validation set $I_{validation}$;
    \State Initialization: the  training, testing and validation set $I_{training}=\emptyset$, $I_{testing}=\emptyset$, $I_{validation}=\emptyset$;
    \Loop {~Randomly sample a user $u$ from $U$:}
        \If{$|I_{training}| \leq 70\% * |I|$ ($|~|$ denotes the size of the set)}
            \State $I_{training} = I_{training} \cup I_{u}$
            \State continue
        \EndIf
        \If {$|I_{testing}| \leq 15\% * |I|$}
            \State $I_{testing} = I_{testing} \cup I_{u}$
            \State continue
        \EndIf
        \State $I_{validation} = I \setminus (I_{testing} \cup I_{training})$
        \State End sampling, quit

    \EndLoop

	\end{algorithmic}
	\label{algo:partition}
\end{algorithm}


\begin{table}[htbp]
  \centering
  \caption{Statistics of datasets after preprocessing.}
    \begin{tabular}{|c|c|c|c|c|}
    \hline
    \textbf{Dataset} & \textbf{users} & \textbf{items} &    \textbf{interaction records} & \textbf{attributes} \\
    \hline
    Yelp & 27,675  & 70,311 & 1,345,606 & 590  \\
    \hline
    LastFM & 1,801  & 7,432 & 76,693 & 33  \\
    \hline
    Kuaishou & 6,861  & 18,612 & 100,699 & 665  \\
    
    \hline
    \end{tabular}%
  \label{t0}%
\end{table}%

\begin{table}[htbp]

  \centering
  \caption{Statistics of the training set, testing set and validation set.}
    \begin{tabular}{|c|c|c|c|c|c|c|}
    \hline
          
          \multicolumn{1}{|c|}{\textbf{Dataset}} & \multicolumn{3}{|c|}{\textbf{users}} & \multicolumn{3}{|c|}{\textbf{interaction records}} \\
    \hline
          & \textbf{Training} & \textbf{Testing} & \textbf{Validation} & \textbf{Training} & \textbf{Testing} & \textbf{Validation}\\
    \hline
    Yelp & 21,309  & 3,205  & 3,161  & 941,928  & 201,845 & 201,833  \\
    
    \hline
    LastFM & 1,260  & 279 & 262  & 53,636  & 11,532 & 11,525 \\
    
    \hline
    Kuaishou & 4,537  & 1,175 & 1,149  & 70,492  & 15,110 & 14,097  \\
    \hline

    \end{tabular}%
  \label{t_new}%
\end{table}%

\subsection{Experimental Setting}\label{ES}
\subsubsection{Baselines}\label{baseline}

While many good conversational recommendation systems such as ~\cite{GeneratingClarifying, AskingClarifying,radlinski-etal-2019-coached, christakopoulou2018q, yu2019visual, priyogi2019preference,sun2018conversational, sardella2019approach, chen-etal-2019-towards}  have been proposed recently, most of them cannot tackle the cold-start scenarios in MCR, hence being incomparable with our ConTS. We try our best to select following representative models as our baselines:

\begin{itemize}
    \item \textbf{Abs Greedy}~\cite{christakopoulou2016towards}: This work trains an offline FM model based on existing users' records. Then it initializes the embedding for the new user with the average of existing users' embeddings. Each time it will recommend the top $k$ items ranked by FM and update the FM according to the rejected item list. But it does not consider any attribute information and recommend without asking any questions. \citet{christakopoulou2016towards} have shown that Abs-Greedy has a better performance than classic Thompson Sampling, so we don't take classic Thompson Sampling as a baseline here.
    
    \item \textbf{EAR}~\cite{lei20estimation}: EAR is a state-of-the-art model on MCR scenario for warm-start users. It trains a policy network by reinforcement learning to decide whether to ask questions or recommend items. The input of the network is an vector which encodes the entropy information of each attribute, user's preference on each attribute, conversation history as well as length of the current candidate list. The output is a probability distribution over all actions, which includes all attributes and a dedicated action for recommendation. It uses an pretrained offline FM model to decide which items to recommend and employs a policy network trained by policy gradient to decide conversational policy. To adapt EAR to our setting, we use the average of existing users' embedding as the embedding of the new user and use the existing user set for offline training. The new user set is used as testing. The rest of the configurations are exactly the same as the original EAR implements. 
    
    \item \textbf{ConUCB}~\cite{zhang2020toward}: Conversational UCB is a recently proposed method to apply bandit to conversational recommendation scenario. It models the attributes and items as different arms and choose them separately by different ranking score. As for the attribute choice policy, the authors put forward four methods in the paper to choose the attribute to ask: Random, Maximal Confidence, Maximal Confidence Reduction and Advanced Strategy. If $n = |{\mathcal{V}_t}|$ denotes the length of current candidate item pool, $k = |{\mathcal{P}_t}|$ denotes the length of the current candidate attribute pool, and $d$ denotes the length of item's embedding, Maximal Confidence Reduction is an $O({n^2 + 2*d^2})$ algorithm and Advanced Strategy is an $O(2* n * d * k + 2*d^2)$ algorithm, while our ConTS is $O(2 * d * (n + k) + d^2)$ (note that $n>>(d+k)$ in our setting). We choose the second method "Maximal Confidence" proposed in the work. The third and the fourth ones in the paper are infeasible in our setting, because a large arm pool in our corpus will result in a gigantic computing complexity. Moreover, all the parameters are tuned following the methods in original paper on validation set. The optimal performance can be achieved when $\lambda=0.8$ and $\tilde \lambda = 1$. We also explore different policy functions in ConUCB and report the best result, which is discussed in details in Section~\ref{unify}. 
\end{itemize}

To validate the key component of our ConTS design, we also compare ConTS with the following variants, each with one component ablated:
\begin{itemize}
    \item \textbf{ConTS-$\textbf{u}_{init}$}: As discussed in Section~\ref{IS}, ConTS
    initializes the mean of new user's preference distribution with the average of existing user embeddings, which is denoted as $\textbf{u}_{init}$. We probe its impact on ConTS by removing it from the model, where we follow~\citet{agrawal2013thompson} to initialize the mean with a zero vector.
    
    \item \textbf{ConTS-$\mathcal{P}_u$}: ConTS models user's currently known preferred attributes $\sum_{p_i\in\mathcal{P}_u}$ when calculating the respected reward for each arm. This is to capture the mutual promotion of items and attributes since they are in the same space. We design experiments to discuss its effect on ConTS by removing $\mathcal{P}_u$ from the reward estimation function (Eq. (\ref{rewards}) in Section~\ref{RA}). As such, the reward estimation exactly follows the contextual Thompson Sampling~\cite{agrawal2013thompson}. The resultant system is denoted as ConTS-$\mathcal{P}_u$. 
    
    \item \textbf{ConTS-exp}: ConTS inherits the sampling mechanism from contextual Thompson Sampling to explore the user's latent unknown preferences. We conduct experiments to investigate the effectiveness of such exploration in ConTS. Specifically, we remove the sampling step and just take the mean of the distribution as user embedding in each turn (we also keep the parameter $\textbf{B}_u$ as identity matrix), aiming to learn the performance of ConTS without exploration.
    
\end{itemize}

\subsubsection{User Simulator}\label{user_simu}
As conversational recommendation is an interactive process, a CRS needs to be trained and evaluated by interacting with users. However, it is expensive and less reproducible to directly recruit human beings to interact with the CRSs. Therefore, we follow the approach of~\citet{lei20estimation, sun2018conversational} to build a user simulator to simulate a user's feedback based on the user's historical user-item interaction records. Specifically, if there is an user-item record $(u,v)$, we treat $v$ as the user's preferred item and its corresponding attributes $\mathcal{P}_v$ as his preferred attributes in this session. We simulate a conversation session for each interaction record and the user will give positive feedback only to item $v$ and attributes $\mathcal{P}_v$ in current session. We also follow~\cite{lei20estimation} to set the max turn $T$ to 15 in our primary experiments but we will also discuss the model performances when setting it to other numbers in Section~\ref{EO}. If the user accept the items recommended by CRS, the conversation will successfully end. If the recommendation is still not successful when achieving the number of max turns, the simulated user will terminate the session due to the lack of patience. Note that we ask enumerated questions on Yelp. If ever user rejects a parent attribute, ConTS will naturally update the algorithm by all the child attributes together. Lastly, it is worth mentioning that we initialize the distribution of embedding for a new user in the first conversation with him. After the first conversation finishes, his parameters are kept and continue to be updated in following conversations.

\subsubsection{Evaluation Metrics}\label{EM}
As what has been done in~\cite{christakopoulou2016towards}, we use metrics in conversational recommendation for evaluation. Following~\citet{lei20estimation, christakopoulou2016towards}, we use two metrics to measure the performance of each model. First we use the success rate (SR@t)~\cite{sun2018conversational} to measure the ratio of successful conversation, i.e., successfully make recommendation by turn $t$\footnote{This means the success rate with at most t turns}. It can be formulated as $SR@t=\frac{\# successful~conversations~by~turn~t}{\# conversations}$. We also report the average turns (AT) needed to end the session. Note that if a conversation session has not been end at turn $T$, we simply set it to be $T$. Higher SR@t represents more successful turns and smaller AT means more efficient conversation. We use one-sample paired t-test to judge the statistical significance.

\subsubsection{Implementation Details}
The working process includes offline and online stages.
In the offline stage, we use the records of all existing users (training set) to train the users, items and attributes embedding using the FM proposed in \cite{lei20estimation}. The training objective is to rank the user's preferred items and attributes higher using the FM proposed in \cite{lei20estimation}. It is optimized using SGD with a regularization strength of 0.001. 

In the online stage, we run the ConTS on the testing set. Specifically, we let our ConTS interact with the new user simulator described in Section \ref{user_simu}. The dialogue sessions are simulated using user--item  historical records simulated in the testing set. The parameters of ConTS are tuned on the simulated conversations from validation set. The values of user's feedback in ConTS is searched in the range of [-8, 8] and we apply grid search to find the optimal settings. We find that the absolute value of positive feedback should be larger while the absolute value of negative feedback should be smaller for better performance. Specifically, the optimal parameters are set as follows:$~{r_{fail\_rec}} = -0.15, {r_{fail\_ask}} = -0.03, {r_{suc\_ask}} = 5,{r_{suc\_rec}} = 5$; besides, the hyperparameters are set as $l=0.01$ and $k=10$. Following EAR \cite{lei20estimation}, the dimension $d$ is set to 64 for embeddings for items, attributes and users.

\subsection{Experiment Results}\label{result}

\begin{table}[htbp]

  \centering
  \caption{Results of the difference on SR@15 and AT of the compared methods for Yelp, LastFM and Kuaishou. Bold scores denote the best in each column, while the underlined scores denote the best baseline. Statistic significant testing is performed ($p<0.01$)}
    \begin{tabular}{|c|c|c|c|c|c|c|}
    \hline
          
          & \multicolumn{2}{|c|}{\textbf{Yelp}} & \multicolumn{2}{|c|}{\textbf{LastFM}} & \multicolumn{2}{|c|}{\textbf{Kuaishou}} \\
    \hline
          & \textbf{SR@15} & \textbf{AT} & \textbf{SR@15} & \textbf{AT} & \textbf{SR@15} & \textbf{AT} \\
    \hline
    Abs Greedy & 0.340  & 14.0115  & 0.0104  & 14.9203 & 0.1395  & 14.0573  \\
    
    \hline
    ConUCB & 0.262  & 13.0117  & \underline{0.1540}  & \underline{14.2279} & 0.1457  & 13.7136  \\
    
    \hline
    EAR & \underline{0.446}  & \underline{11.5825}  & 0.0640  & 14.3278 & \underline{0.2362}  & \underline{13.4753}  \\
    \hline
    \hline

    ConTS-$\textbf{u}_{init}$ & 0.734  & 8.6195  & 0.1514  & 13.9767& 0.4644  & 12.0534  \\
    
    \hline
    ConTS-${\mathcal{P}_u}$ & 0.328  & 12.8060  & 0.1505  & 14.0210 & 0.4031  & 12.2949  \\
    
    \hline
    ConTS-exp & 0.491  & 10.9975  & 0.0610  & 14.5912 & 0.1446  & 13.7879  \\
    
    \hline
    ConTS   & \textbf{0.756}* & \textbf{8.5808}* & \textbf{0.1628}* & \textbf{13.9162}* & \textbf{0.4872}* & \textbf{11.8703}*\\
    \hline
    \end{tabular}%
  \label{t1}%
\end{table}%

\begin{figure*}[b]
\centering
\subfigure[]{
\includegraphics[height=6.1cm,width=9cm]{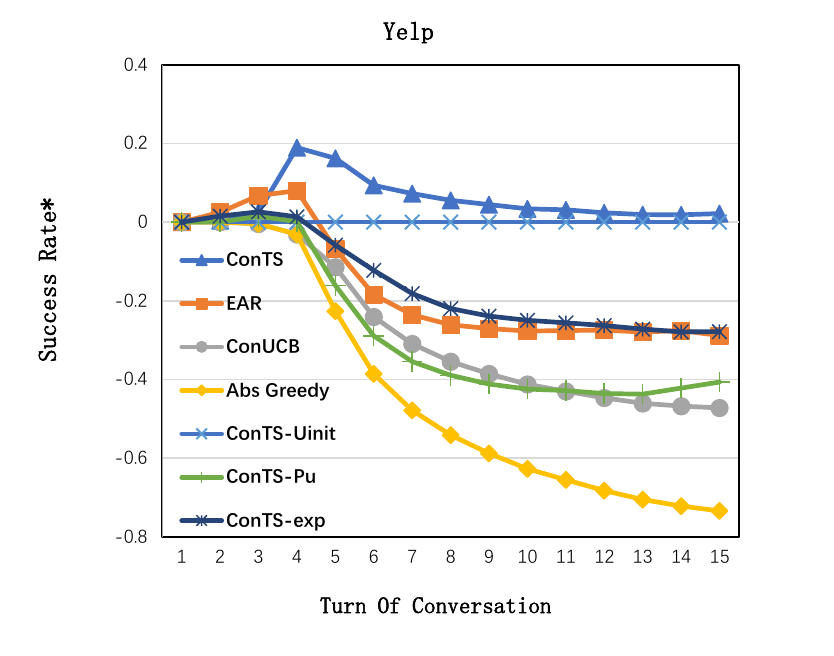}
}

\subfigure[]{
\includegraphics[height=6.2cm,width=8.7cm]{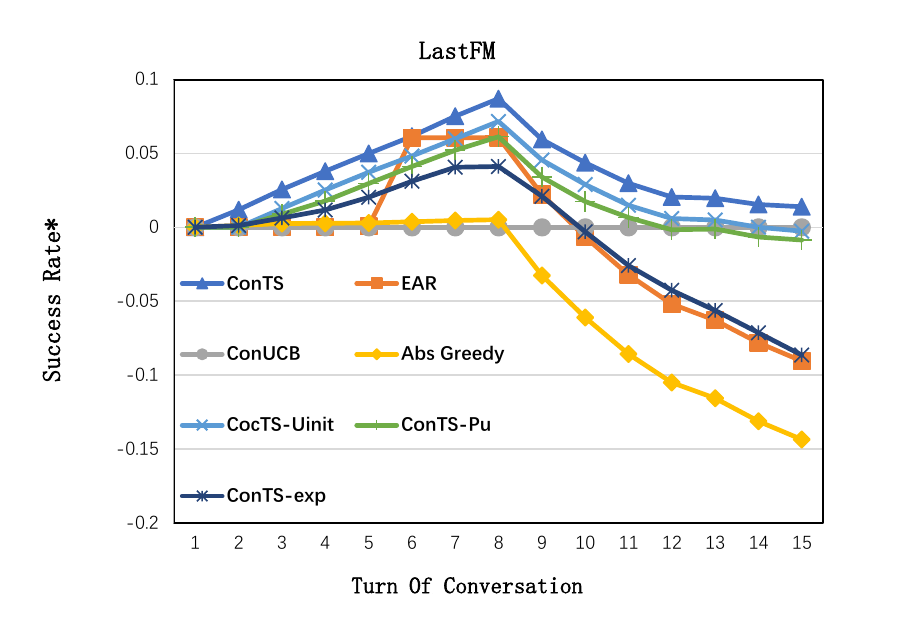}
}

\end{figure*}
\addtocounter{figure}{-1}
\begin{figure*}
\addtocounter{figure}{1} 
\centering
\subfigure[]{
\includegraphics[height=6.0cm,width=8.5cm]{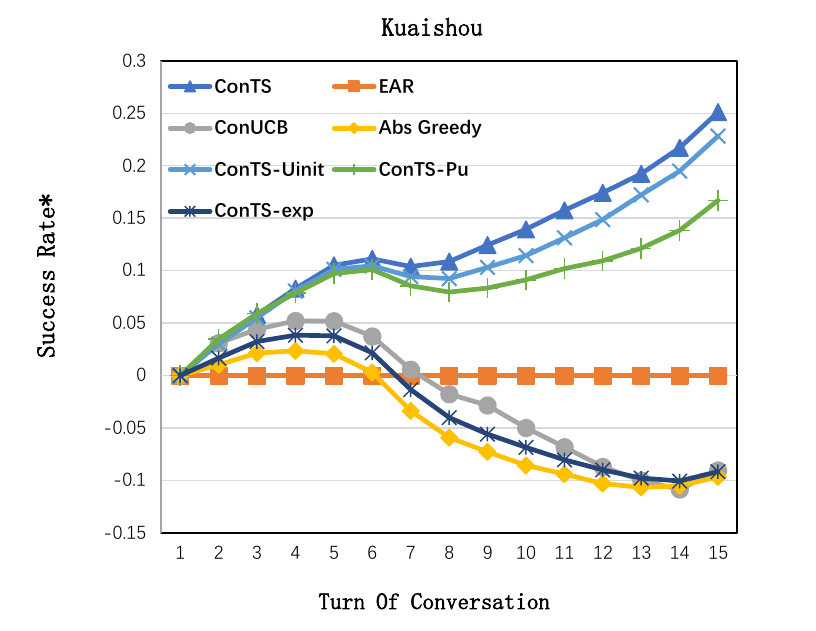}
}
\caption{Success Rate* (SR@t) of compared methods at different conversation turns on Yelp, LastFM and Kuaishou. Success Rate* denotes the difference of SR between each method and the system represented by the line of $y=0$. We set ConTS-$\textbf{u}_{init}$ as $y=0$ on Yelp, ConUCB as $y=0$ on LastFM  and EAR as $y=0$ on Kuaishou dataset to facilitate visualization.}
\label{f3}
\end{figure*}

Figure~\ref{f3} and Table~\ref{t1} report the overall experiment of results. Specifically, Figure~\ref{f3} compares the recommendation Success Rate (SR)@t at each turn ($t$=1 to 15) of different models. We set ConTS -$\textbf{u}_{init}$ (i.e, ConTS without $\textbf{u}_{init}$) as $y=0$ on Yelp, ConUCB as $y=0$ on LastFM and EAR as $y=0$ on Kuaishou dataset for a better view. 
From Figure~\ref{f3}, we can see that ConTS achieves a higher success rate in nearly all turns than the other models with the exception of the first several turns on Yelp Dataset. We find it is because the policy is a bit random at the very beginning of each conversation session, hence the performances are not very indicative. This is especially severe on the datasets with a larger item pool like Yelp. Table~\ref{t1} reports the performance of all models on average turn (AT) and final success rate SR@15 (we set the max turn to be 15 here). It is clear to see that ConTS significantly outperforms all the other models on both (SR)@15 and AT. We notice that nearly all the mentioned models perform much worse on LastFM and Kuaishou than on Yelp.
The main reason is that ConTS asks enumerated questions on Yelp ({\it c.f.} Section~\ref{DS}) where the user chooses the child-level attributes which is finer grained (a numerated question asks more than 20 original attributes on average in a turn). This makes the candidate items shrink much faster than in LastFM and Kuaishou corpus which adopts binary question setting. This phenomenon has also been discussed by \citet{lei20estimation} in detailed. 

The experiments also { demonstrate the effectiveness of asking for attribute preference}. We can see that Abs-Greedy, the only model unable to ask attributes, almost has the worst success rate on all turns. We argue that asking attributes significantly help with this scenario on two points: (1) eliminating unqualified candidate items, i.e., the items that do not contains all the user preferred attributes; and (2) helping better estimating user preference, i.e., estimating arm's reward. While the first point is obvious, the second point can be validated by comparing ConTS and ConTS-${\mathcal{P}_u}$. 
The only difference between ConTS and ConTS-${\mathcal{P}_u}$ is that ConTS models user's known preferred attributes $\sum_{p_i\in\mathcal{P}_u}$ into calculation of arms' reward. From Figure~\ref{f3} and Table~\ref{t1}, we can clearly see ConTS-${\mathcal{P}_u}$ significantly underperforms ConTS. { The results suggests that calculating rewards with $\sum_{p_i\in\mathcal{P}_u}$ contributes significantly to estimating user's preference in ConTS.}

{By comparing EAR with ConTS, we can see the importance of exploration for cold-start users.}
EAR is the state-of-the-art model for warm start users in the MCR scenario~\cite{lei20estimation}. It uses the same function to estimate the user preference (Eq. (\ref{rewards}) in Section~\ref{RA}). The major difference between ConTS and EAR is that ConTS customizes contextual Thompson Sampling to make exploration-exploitation balance while EAR uses reinforcement learning to only exploit the learned user's preference in the training stage. The lack of the exploration mechanism leads to the inferior performance compared with our ConTS. The importance of exploration can further be demonstrated by comparing ConTS with ConTS-exp which is the variant of our ConTS by only removing the exploration mechanism in contextual Thompson Sampling. From Figure~\ref{f3} and Table~\ref{t1}, we can observe that ConS-exp has worse performance than that of ConTS.

{We have also studied the impact of initialization.} ConTS initializes the distribution of the new user's embedding by setting the initial mean as the average of the embeddings of existing users (i.e., $\textbf{u}_{init}$ in Eq.(\ref{init})). We argue this step is also useful. The superior performance of ConTS to ConTS-$\textbf{u}_{init}$ validates this ({\it c.f.} Figure~\ref{f3} and Table~\ref{t1}). Interestingly, we also notice the success rate curve on Yelp in Figure~\ref{f3} of ConTS-$\textbf{u}_{init}$ gradually approaches to the curve of ConTS as the number of conversation turns increases. It is due to the updates. As the conversation evolves, the CRS gets more information of the user preference from his feedback. Accordingly, the model gets less benefits from the initialization. This trend is not obvious on LastFM and Kuaishou, we believe that is due to the intrinsically different question settings. The binary question adopted in the LastFM and Kuaishou dataset helps less than the enumerated question in decreasing the candidates pool, making the candidate pool much larger. As a result, the updates are less efficient.

\subsection{Discussion: Unifying Attributes and Items}\label{unify}

The key contribution in our ConTS is the holistic modeling: we seamlessly unify the attributes and items in the same arm space, naturally fitting them into the framework of contextual Thompson Sampling. The model only needs to calculate reward for each arm using a unified function (Eq. (\ref{rewards})). This reduces the \emph{conversation policy questions} (e.g., deciding what attributes to ask, what items to recommend and whether to ask attributes or making recommendations) as a single problem of arm choosing. This also helps to capture the mutual promotion between items and attributes by conside ring both in Eq. (\ref{rewards}). In this section, we discuss such issues by investigating ConUCB deeply, which also takes the items and attributes as arms in the framework of bandit algorithm but models them separately. We report the experimental results on Yelp dataset, upon which the original ConUCB~\cite{zhang2020toward} is built, as they are more representative. Conclusions in this section also apply to the rest two datasets.




Specifically, in the original paper~\cite{zhang2020toward} of ConUCB, the model decides when to ask attributes based on the heuristics that a system only asks attributes if $b(t)-b(t-1)>0$, where $b(t)$ is a function of the turn number in conversation. However, different function may work better in different datasets or application scenario. This makes the model be less robust as it is hard to choose a universal rule for all cases. For example, the author tried several different functions to control the policy, such as $5 * \log \left\lfloor t \right\rfloor $, $10 * \log \left\lfloor t \right\rfloor $,$15 * \log \left\lfloor t \right\rfloor $, ${\text{10}} * \log \left\lfloor {\frac{t}{{50}}} \right\rfloor $,ect. The paper compares the model's performance on those different functions, but it does not give an compelling strategy about how to choose it according to different situations. In the original paper~\cite{zhang2020toward}, the authors make a discussion that the model may perform better on Yelp dataset if we choose the function which encourages asking more questions. However, to make sure we can make successful recommendation in limited conversation turns, we should always find out the balanced point between asking questions and recommending items. If we follow the way in ConUCB and luckily seek out a suitable function with good performance on the current dataset from countless candidate functions, we can never make sure it will work well next time when we face a different recommendation situation.
To investigate the robustness of the function, we compare the performance of ConUCB by choosing different functions on Yelp. Table~\ref{t2} reports the results. We can see that, {the performance fluctuates hugely with regard to different function. Different from ConUCB, our ConTS is totally data driven -- it learns conversation policies totally based on the model which is estimated based on user feedback, being more intelligent and portable to new application scenarios.}

\begin{table}[htbp]

  \centering
  \caption{SR@15 and AT of ConUCB w.r.t. different functions to decide \emph{whether to ask attributes} on Yelp is statistically significant ($p<0.01$).}
    \begin{tabular}{|c|c|c|}
    \hline
    \textbf{Policy Function $b(t)$} & \textbf{SR@15} & \textbf{AT} \\
    \hline
    $5 * \left\lfloor {\log (t)} \right\rfloor $ & 0.117  & 13.8065  \\
    \hline
    $5 *{\log (t)}  $ & 0.119  & 13.7725  \\
    \hline
    $10 *{\log (t)}  $ & 0.262  & 13.0117  \\
    \hline
    $15 *{\log (t)}  $ & 0.162  & 12.8927  \\
    
    \hline
    \end{tabular}%
  \label{t2}%
\end{table}%

We also investigate the mutual promotion of the items and attributes. In ConUCB, the model uses two separate methods to choose the items and attributes where the two are totally different. We probe the mutual promotion by replacing the attribute choosing function (i.e., the \emph{Maximal Confidence}) used in ConUCB with our reward estimation function\footnote{It is not portable to experiment with the item choosing function as it requires complex mathematics for adaptation. Hence, we only experiment with the attribute choosing function in ConUCB.} (Eq. (\ref{rewards})), which is a modified FM. Namely:

\begin{equation}
\begin{aligned}
   \bbE[r(a,u,\mathcal{P}_u)]=  \bu^{\mt} \mathbf{x}_{a}
    + \sum_{p_i\in\mathcal{P}_u} {\mathbf{x}_a}^T \mathbf{p}_i .
\end{aligned} 
\end{equation}

\begin{table}[htbp]

  \centering
  \caption{Comparing Maximal Confidence and Modified FM as attribute choosing policy in ConUCB on Yelp. Statistic significant testing is performed ($p<0.01$).}
    \begin{tabular}{|c|c|c|c|c|}
    \hline
          
          & \multicolumn{2}{|c|}{\textbf{Maximal Confidence}} & \multicolumn{2}{|c|}{\textbf{Modified FM}} \\
    \hline
          \textbf{Policy $b(t)$}& \textbf{SR@15} & \textbf{AT} & \textbf{SR@15} & \textbf{AT} \\
    \hline
    $5 * \left\lfloor {\log (t)} \right\rfloor $ & 0.117  & 13.8060  & 0.125  & 13.7007  \\
    
    \hline
    $5 *{\log (t)}  $ & 0.119  & 13.7725  & 0.532  & 10.3027  \\
  
    \hline
    $15 *{\log (t)}  $ & 0.197  & 13.0117  & 0.598  & 9.7536  \\
    
    \hline
    \end{tabular}%
  \label{t3}%
\end{table}%

Table~\ref{t3} shows that (Eq.~\ref{update4}) helps to improve the performance of ConUCB by a large degree. { This demonstrates the effectiveness of the mutual promotion between attributes and items, thus further validating our idea of unifying the modeling of attributes and items in the conversational recommendation scenario. }

\subsection{Discussion: Thompson Sampling v.s. UCB}

Despite the difference between separately or jointly modeling item and attributes, ConTS and ConUCB also differ in the bandit algorithm framework they adopt: ConTS utilizes Thompson Sampling while ConUCB utilizes UCB. To further discuss the effect of our strategy of unifying attributes and items modeling, we  design another experiment to remove the effects of different bandit algorithms. To do that, we adapt LinUCB~\cite{li2010contextual} to our setting by replacing the sampling step with an upper confidence bound (all the rest modules in ConTS remain the same). The reward for each arm in the adapted LinUCB (denoted as Seamless-UCB) is calculated as follows:

\begin{equation}
\begin{aligned}
    \bbE[r(a,u,\mathcal{P}_u)]= \mathbf{u}^T\mathbf{x_{a}} 
    + \sum_{p_i\in\mathcal{P}_u} \mathbf{x_{a}}^T \mathbf{p}_i
    + \alpha \sqrt{{\mathbf{x}_a}^T{\textbf{A}_u^{ - 1}}\mathbf{x}_a}~,
\end{aligned} 
\end{equation}
where $\alpha \sqrt{{\mathbf{x}_a}^T{\textbf{A}_u^{ - 1}}\mathbf{x}_a}$ denotes the upper confidence bound and $\mathbf{u}$ and $\textbf{A}_u$ are updated as:
\begin{equation}
\begin{aligned}
   \textbf{A}_u =  \textbf{A}_u + \textbf{x}_{a(t)}{\textbf{x}_{a(t)}}^T
\end{aligned} 
\end{equation}
\begin{equation}
\begin{aligned}
   r_a' = r_a - {\textbf{x}_{a(t)}}^T (\textbf{u}_{init} + \sum_{p_i\in\mathcal{P}_u} \mathbf{p}_i)
\end{aligned} 
\end{equation}
\begin{equation}
\begin{aligned}
   \textbf{b}_u = \textbf{b}_u + r_a' * \mathbf{x}_a
\end{aligned} 
\end{equation}
\begin{equation}
\begin{aligned}
   \textbf{u} = \textbf{A}_u^{ - 1}{\textbf{b}_u}.
\end{aligned} 
\end{equation}

From Table~\ref{t4} we can { see that the Seamless-UCB, which uses our unifying item and attribute strategy in the UCB framework, still outperforms ConUCB. This also demonstrates the advantage of our seamless modeling strategy. }


Interestingly, from the comparison of Seamless-UCB and ConTS, we can see that {Thompson Sampling has a better performance in our setting.} This is inline with previous research~\cite{russo2014learning, osband2017optimistic} which reports Thompson Sampling performs better on many real-world scenarios and also reaches a smaller theoretical regret bound. 
This interesting discovery indicates that Thompson Sampling might be more powerful than UCB in conversational recommendation scenario. However, we leave more discussions to the future works as the comparison between Thompson Sampling and UCB is still an open question~\cite{russo2018tutorial}.

\begin{table}[htbp]

  \centering
  \caption{Comparison between ConTS, ConUCB and Seamless-UCB. Bold scores are the best in each column. Statistic significant testing is performed ($p<0.01$).}
    \begin{tabular}{|c|c|c|c|c|c|c|}
    \hline
          
          & \multicolumn{2}{|c|}{\textbf{Yelp}} & \multicolumn{2}{|c|}{\textbf{LastFM}} & \multicolumn{2}{|c|}{\textbf{Kuaishou}} \\
    \hline
          \textbf{}& \textbf{SR@15} & \textbf{AT} & \textbf{SR@15} & \textbf{AT} & \textbf{SR@15} & \textbf{AT} \\
    \hline
    
    ConUCB & 0.2620  & 13.0117  & 0.1540  & 14.2279 & 0.1457  & 13.7136  \\
    
    \hline
    Seamless-UCB & 0.7394  & 8.7785  & 0.1593  & 14.0126 & 0.4136  & 12.7126 \\
  
    \hline
    ConTS   & \textbf{0.7558}* & \textbf{8.5808}* & \textbf{0.1628}* & \textbf{13.9162}* & \textbf{0.4872}* & \textbf{12.1033}*\\

    \hline
    \end{tabular}%
  \label{t4}%
\end{table}%

\subsection{Discussion: Effect of Max Conversation Turn}\label{EO}
 As mentioned in Section~\ref{EM}, we set the max conversation turn to 15 in the above experiments. However, some real-world scenarios may acquire smaller max conversation turn due to the less patience of users. In this section, we conduct experiments to explore the situation with a much tighter cap on utterance. Specifically, we compare the performance of all mentioned methods by setting two smaller max conversation turn (\textbf{7} and \textbf{10}) on Yelp, LastFM and Kuaishou. 
Here, we report the results when max conversation turn 
is \textbf{7} instead of \textbf{5}. The reason is that several models, such as EAR, ConUCB, Abs Greedy, have a terrible performance on the first five turns (in Figure~\ref{f3}). It is hard to get meaningful interpretations. Therefore, we report the results in turn 7 which can better reveal the difference between these compared methods.

\begin{table}[htbp]

  \centering
  \caption{Results of difference on SR@7 and AT of compared methods when the max conversation turn is \textbf{7}. Bold scores denote the best in each column, while the underlined scores denote the best baseline. Statistic significant testing is performed ($p<0.01$). }
    \begin{tabular}{|c|c|c|c|c|c|c|}
    \hline
          
          & \multicolumn{2}{|c|}{\textbf{Yelp}} & \multicolumn{2}{|c|}{\textbf{LastFM}} & \multicolumn{2}{|c|}{\textbf{Kuaishou}} \\
    \hline
          & \textbf{SR@7} & \textbf{AT} & \textbf{SR@7} & \textbf{AT} & \textbf{SR@7} & \textbf{AT} \\
    \hline
    Abs Greedy & 0.1798  & 6.6899  & 0.0046  & 6.983 & 0.0637  & 6.7793  \\
    
    \hline
    ConUCB & 0.1692  & 6.5184  & 0.0024  & 6.9876 & \underline{0.1029}  & \underline{6.6072}  \\
    
    \hline
    EAR & \underline{0.2425}  & \underline{6.2305}  & \underline{0.0606}  & \underline{6.9215} & 0.0973  & 6.8606  \\
    
    \hline
    \hline
    
    ConTS-$\textbf{u}_{init}$ & 0.4784  & 5.9671  & 0.0600  & 6.8282 & 0.1915  & 6.3848  \\
    
    \hline
    ConTS-${\mathcal{P}_u}$ & 0.1237  & 6.5986  & 0.0522  & 6.8612 & 0.1828  & 6.3893  \\
    
    \hline
    ConTS-exp & 0.2967  & 6.2155  & 0.0406  & 6.8979 & 0.0838  & 6.8746  \\
    
    \hline
    ConTS   & \textbf{0.5508}* & \textbf{5.3995}* & \textbf{0.0752}* & \textbf{6.7512}* & \textbf{0.2012}* & \textbf{6.3610}*\\
    \hline
    \end{tabular}%
  \label{t6}%
\end{table}%

\begin{table}[htbp]

  \centering
  \caption{Results of difference on SR@10 and AT of compared methods when the max conversation turn is \textbf{10}.Bold scores denote the best in each column, while the underlined scores denote the best baseline. Statistic significant testing is performed ($p<0.01$). }
    \begin{tabular}{|c|c|c|c|c|c|c|}
    \hline
          
          & \multicolumn{2}{|c|}{\textbf{Yelp}} & \multicolumn{2}{|c|}{\textbf{LastFM}} & \multicolumn{2}{|c|}{\textbf{Kuaishou}} \\
    \hline
          & \textbf{SR@10} & \textbf{AT} & \textbf{SR@10} & \textbf{AT} & \textbf{SR@10} & \textbf{AT} \\
    \hline
    Abs Greedy & 0.2569  & 9.1584  & 0.0065  & 9.9649 & 0.0918  & 9.5315  \\
    
    \hline
    ConUCB & 0.2145  & 8.7965  & 0.0275  & 9.9228 & 0.1276  & \underline{9.2240}  \\
    
    \hline
    EAR & \underline{0.3502}  & \underline{8.2744}  & \underline{0.0619}  & \underline{9.6966} & \underline{0.1775}  & 9.3712 \\
    
    \hline
    \hline
    
    ConTS-$\textbf{u}_{init}$ & 0.6269  & 7.1572  & 0.0964  & 9.5764 & 0.2918  & 8.6071  \\
    
    \hline
    ConTS-${\mathcal{P}_u}$ & 0.2023  & 9.0030  & 0.0849  & 9.6435 & 0.2687  & 8.6696  \\
    
    \hline
    ConTS-exp & 0.3773  & 8.1632  & 0.0645  & 9.7279 & 0.1089  & 9.3899  \\
    
    \hline
    ConTS   & \textbf{0.6609}* & \textbf{6.4655}* & \textbf{0.1112}* & \textbf{9.4549}* & \textbf{0.3168}* & \textbf{8.5216}*\\
    \hline
    \end{tabular}%
  \label{t7}%
\end{table}%

Table~\ref{t6} and Table~\ref{t7} report the results when the max turn set to \textbf{7} and \textbf{10} separately. It is clear to see that ConTS still significantly outperforms all other models on both SR@(7) (or SR@10) and AT. This demonstrates that ConTS has a better performance even when given a much smaller max conversation turn. 

Apart from it, it is interesting to see the gap of performance between ConTS and other methods becomes larger when the max conversation turn is bigger. We attribute it to two reasons: First, since we make recommendation for cold-start users, all the policies are somehow random at the beginning of each conversation session, which makes the difference not obvious when the max conversation turn is small; Second, ConTS is able to better capture user's preference in each conversation turn than the other methods, thus making the accumulated advantage larger as the number of conversation turn increases.

\subsection{Discussion: Effects of Cold-Start Degrees}

In the above discussions, we have explored the cold start cases where historical user-item interaction records are totally unavailable for the new users. That is the typical scenario which the Thompson Sampling is originally designed for ~\cite{graepel2010web, granmo2010solving, may2011simulation, murphy2012machine, agrawal2013thompson}. However, in real applications, another common cold-start case is that limited historical records are available. There comes a natural question: what is the performance of ConTS in such case? 

To have more idea of this question, we test the performance of ConTS in different cold-start degrees. Specifically, we randomly delete the user-item interaction records with the ratio of $\theta$ for each testing user and add such deleted records into the training and validation sets\footnote{we follow 7:1.5 when distributing such records.}. Therefore, $\theta$ can be used to quantify the ''cold start degree''. Then, we use the new training data to train the offline FM model discussed in Section~\ref{RA} and directly obtain the initial embeddings for testing users for the online updating\footnote{Comparing to the case where no historical interactions for the testing user and we average the embedding for all existing users as the embedding for the testing user}. To make the discussion more interesting, we compare the performance of EAR under such different cold-start degrees\footnote{Note that EAR and ConTS use the same method (i.e., Eq.\ref{rewards}) to rank items. The difference is that ConTS uses Thompson Sampling to achieve EE balance for cold-start interaction strategy while EAR employs policy network trained by reinforcement learning for warm-start interaction strategy.} For EAR, we follow~\cite{lei20estimation} to train the FM model on the training set and train the policy network on the validation set.

\begin{figure}
\begin{minipage}[t]{1\linewidth}
    \includegraphics[width=\linewidth]{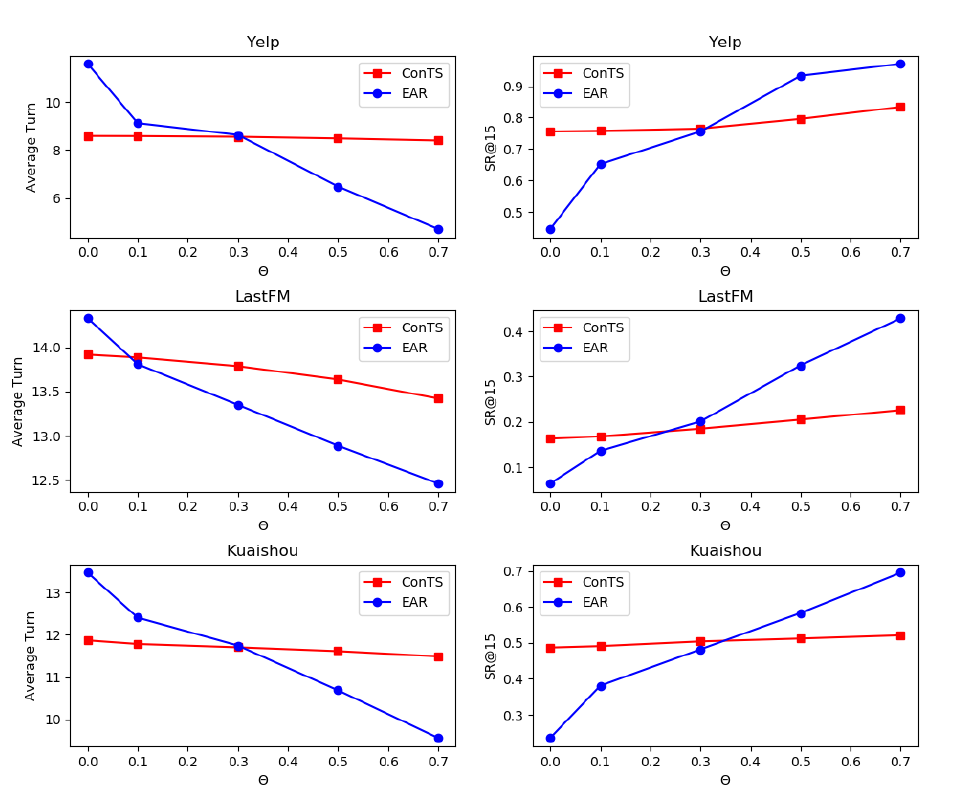}
\end{minipage}%

\caption{SR@15 and AT of EAR and ConTS on Yelp, LastFM and Kuaishou with different cold-start degrees. $\theta$ quantifies the ``cold start degree".}
\label{new_exp}
\end{figure}

The experimental results are demonstrated in Figure~\ref{new_exp}.  The trend is consistent on all three datasets. When there are very few existing historical records (e.g., $\theta < 0.1$), ConTS consistently gets better performances on both Average Turn and SR@15. As the existing records become more, the performance of EAR increases much sharper than ConTS and eventually outperforms ConTS by a large margin. 
This suggests that in the real application, it is better to apply ConTS to the users when there are very few historical records. When the records become more, it would be more helpful to shift to warm start solutions like EAR. However, to the best of our knowledge, few research has conduct conclusive discussions on the comparison between the bandit-based and warm-start recommendation systems in terms of the degree of cold start. We hope our discussions can be a start to trigger more explorations in the future.


\subsection{Case Study}

\begin{figure*}
\centering
\subfigure[The user likes a video about a famous beautiful female tennis player.]{
\includegraphics[height=5.5cm,width=12cm]{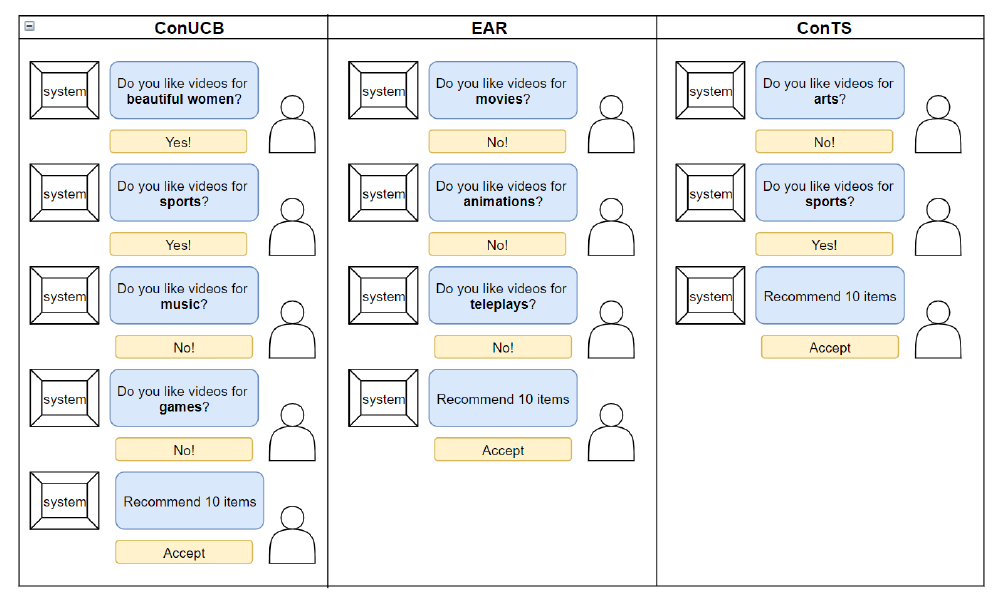}}
\subfigure[The user likes a video about the development of Buddhism in China.]{
\includegraphics[height=5.5cm,width=12cm]{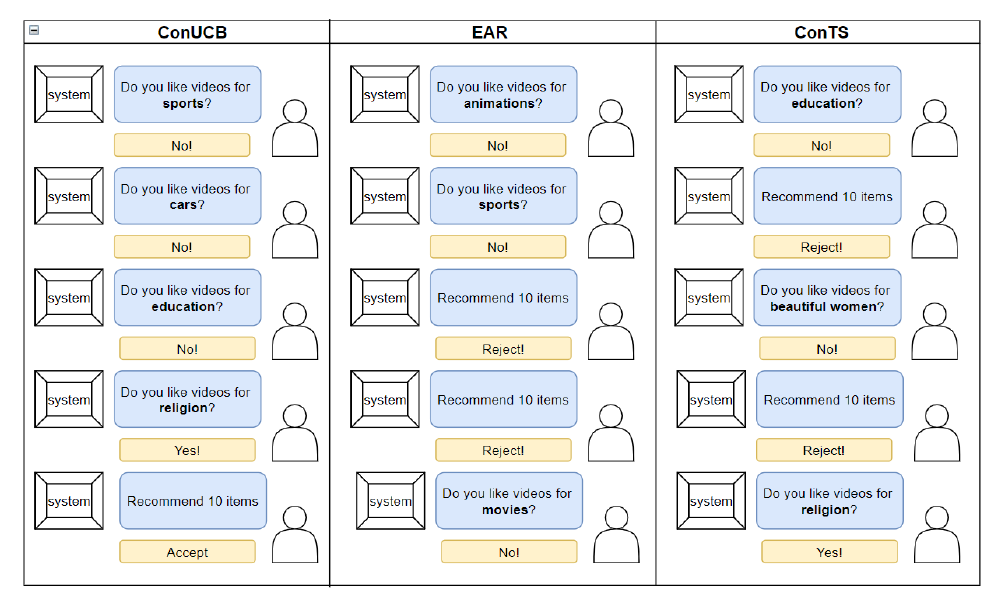}}
\subfigure[The user likes a video about a recently released movie.]{
\includegraphics[height=5.5cm,width=12cm]{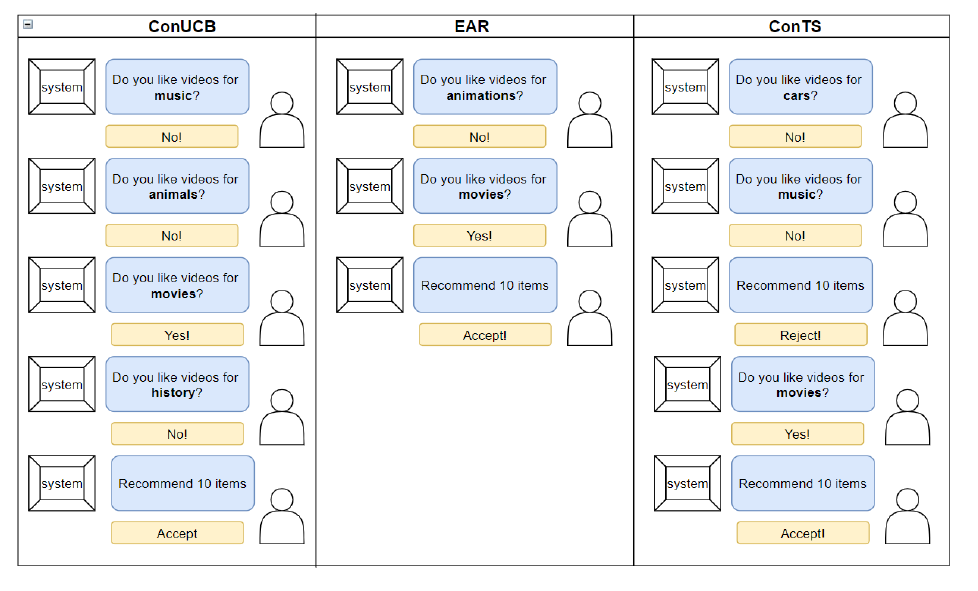}}
\caption{Samples of conversation of ConUCB(left) , EAR(middle) and ConTS(right) of three users.}
\label{case}
\end{figure*}

In this section, we perform three case studies (in Figure~\ref{case}) on real interaction records of users in Kuaishou dataset for one-attribute questions by two baselines ConUCB, EAR and our proposed model ConTS. In case (a), the user wants a video of a famous beautiful female tennis player. The video has two attributes: \emph{beautiful women} and \emph{sports}. As mentioned in Section~\ref{unify},  ConUCB implements a handcrafted way to decide whether to ask or recommend. From Figure~\ref{case} (a), we can see that ConUCB keeps asking attributes for the first four turns in the conversation. However, we notice that the system has already known all the user's preferred attributes after the first two turns. It is clearly the perfect time to recommend items as asking more questions will not bring additional benefits. This reveals less flexibility of ConUCB in making conversation policies. EAR decides the conversation policy questions based on historical interaction records of existing users by a pre-trained FM model a policy network. As a result, it tends to ask some popular attributes to new users according to training data, failing to explore new user's interest. For example, in the conversation in Figure~\ref{case}(a), EAR insists asking attributes with similar themes (such as "movies", "animations" and "teleplays"). Although these attributes are popular among existing users, the new user may not be interested in them at all. In contrast, ConTS can explore new user's unknown interest. In the conversation, since the user gives negative feedback to attribute "arts", ConTS learns to explore his interest on some different attributes, like "sports". And once it captures user's preference on an attribute, it directly recommends items aiming to successfully end the conversation within short conversation turns.

Apart from the first case where ConTS get better performances than the baselines, we also give two more "bad" cases where ConTS is not the best. In Figure~\ref{case}(b), only ConUCB successfully completes recommendation within the first five conversation turns. The user in this case prefers to watch a video about the development of Buddhism in China. It has the attribute \emph{religion}. All the three methods have difficulties in inferring the user's interest at the beginning and the user gives negative feedback to the first questions from all the three methods. However, ConUCB keeps asking questions in the first four turns as its rule-based policy function force it to do so, while EAR and ConTS have less patience and turn to recommending items when the user gives negative feedback. This indicates that ConUCB may have better performances when the user's interest is hard to infer, because we can employ a rule-based policy function to force it to ask more questions before making recommendations. This observation is in line with the results in Table~\ref{t1}: ConUCB gets better relative performance on LastFM which is supposed to be the most difficult dataset (\emph{c.f.} Table~\ref{t1} where all models consistently get worst performances in LastFM dataset). Although it is possible for EAR and ConTS to learn to patiently ask more questions before making recommendations, the process is less tractable the policies are trained automatically.

The last case (Figure~\ref{case}(c)) demonstrates a user who likes a video introducing a recently released movie. It has the attribute \emph{movie}. As introduced in the first paragraph of this section, movie is a quite popular attribute among the existing users. Since EAR tends to ask popular attributes to new users, it quickly hits the user's preferred attribute and then successfully recommend items within three conversation turns. In contrast, ConTS and EAR tend to further explore user's interest on some attributes which are not very popular, such as "cars" or "history". This indicates that if the new user has similar preference with the average of existing users', EAR can be a more efficient method.

\section{conclusion and future works}

In summary, this paper focuses on the cold-start problems in conversational recommendation. We customized contextual Thompson Sampling, a classic bandit algorithm, to conversation scenario, resulting in our ConTS model. ConTS makes the key contribution to seamlessly unify items and attributes as undifferentiated arms in one space. As such, our ConTS addresses the three \emph{conversation policy questions} --  what attributes to ask, what items to recommend, whether to ask attributes or recommend items in a turn-- as a single problem of arm choosing based on a unified reward estimation function. With this simple strategy, ConTS nicely fits the conversational recommendation problem in the framework of contextual Thompson Sampling, achieving good EE balance and intelligent dialogue policy within a holistic model.



We designed a series of experiments on Yelp, LastFM and Kuaishou datasets and empirically validated the effectiveness of our proposed ConTS on various settings. First, the model outperforms various strong baselines (e.g., Abs Greedy, EAR and ConUCB) both on success rate in nearly all turns and average turn for successful recommendation. Second, we validated important components, i.e., initialization, reward estimation and exploration mechanism, by ablating each of them from our ConTS. The results demonstrate that each of the components do help to improve ConTS's performance. Then, we discussed ConUCB in details as it also uses the bandit algorithm to model item and attributes together but handle them separately. We found that, separately modeling items and attributes cannot perform stably and hard to capture the mutual promotion of both attributes and items. This further demonstrates the advantages of our strategy of seamlessly unifying items and attributes in the same space. Furthermore, we conducted experiments by setting a smaller max conversation turn and found that our ConTS still performs well with a tighter cap on conversation. We also did a case study to compare the performance of different methods.

There are still works to do in the future to explore conversational recommendation in cold-start scenario. First, we can make improvement on bandit algorithms by exploring different reward estimation functions and exploring more update strategies. Second, we can also consider more complex problem settings. For example, we can consider how to handle "don't know" of "don't care" responses. Other types of response are also possible, for example "I'd prefer X to Y", and some systems would be expected to handle this (especially if they presented in natural language). Flexibly handling this sort of response would be a very practical advance. Third, we can extend the current framework into neural fashion, leveraging on the powerful modeling capability of neural network to model more complex patterns of user preferences.

\section{supplementary}\label{supp}

In this supplementary, we will try to illustrate how contextual Thompson Sampling help to keep EE balance in an intuitive way. As shown in Algorithm 1, we use a multidimensional Gaussian distribution to describe each user's preference on arms. So the mean $\bm{\mu}_u$ and covariance ${l^2}\textbf{B}_u^{ - 1}$ determine the characters of the distribution. Each time when the agent plays an arm $a(t)$ with embedding ${\textbf{x}_{a(t)}}$ and gets the feedback $r$, it will firstly update the covariance ${l^2}\textbf{B}_u^{ - 1}$ by the formula:
\begin{equation}
\begin{aligned}
  \mathbf{B}_u =  \mathbf{B}_u + \mathbf{x}_{a(t)}{\mathbf{x}_{a(t)}}^T .
\end{aligned} 
\end{equation}

We focus on the diagonal elements of $\textbf{B}_u$ which have the biggest impact on the sampling result. We denote these elements as a vector ${\bm{\lambda} _{\text{B}}}$, so they will be updated like this:
\begin{equation}
\begin{aligned}
{\bm{\lambda} _{B}=\bm{\lambda} _{B}+{\mathbf{x}_{a(t)}^{2}}}
\end{aligned} 
\end{equation}

Apparently, the diagonal elements of matrix $\textbf{B}_u$ will increase after taking each action (asking attribute or recommending item). Since the covariance matrix is ${l^2}\textbf{B}_u^{ - 1}$, the diagonal elements of it will decrease instead. This means our uncertainty on estimating the user's preference on the arms also reduces. Next time when we do sampling again, the result will be closer to the mean value $\bm{\mu}_u$ for those played arms because of the smaller covariance value due to the decreased 
values in covariance matrix. It tells the system to do less exploration on those arms since we have already know the user's preference on them. 

As for the mean $\bm{\mu}_u$, we update it in this way:
\begin{equation}
\begin{aligned}
   \textbf{B}_u =  \textbf{B}_u + \textbf{x}_{a(t)}{\textbf{x}_{a(t)}}^T
\end{aligned} 
\end{equation}
\begin{equation}
\begin{aligned}
   {\emph{\textbf{f}}_u} = {\emph{\textbf{f}}_u} + r_a' * {\textbf{x}_{a(t)}}
\end{aligned} 
\end{equation}
\begin{equation}
\begin{aligned}
   {\bm{\mu} _u} = \textbf{B}_u^{ - 1}{\emph{\textbf{f}}_u} .
\end{aligned} 
\end{equation}

Consider the formula separately. First we fix $\textbf{B}_u$ and investigate the impact of ${\emph{\textbf{f}}_u}$. In that case, after getting the feedback of arm $a(t)$ and updating ${\bm{\mu} _u}$, the agent will calculate the reward of the played arm $a(t)$ next time as follows. The expectation of the new estimated reward is:
\begin{equation}
\begin{aligned}
\bbE[rewar{d_{new}}]&={\textbf{x}_{a(t)}}^T{\bm{\mu} _{new}}\\ 
&={\textbf{x}_{a(t)}}^T{\textbf{B}_u^{ - 1}}{\emph{\textbf{f}}_{new}}\\
&={\textbf{x}_{a(t)}}^T{\textbf{B}_u^{ - 1}}({\emph{\textbf{f}}} + {\textbf{x}_{a(t)}} * r_a')\\ 
&={\textbf{x}_{a(t)}}^T{\textbf{B}_u^{ - 1}}{{\emph{\textbf{f}}}_{old}} + {\textbf{x}_{a(t)}}^T{\textbf{B}_u^{ - 1}}{\textbf{x}_{a(t)}} * r_a'\\ 
&=\bbE[rewar{d_{old}}] + {\textbf{x}_{a(t)}}^T{\textbf{B}_u^{ - 1}}{\textbf{x}_{a(t)}} * r_a'  .
\end{aligned}
\end{equation}

Since ${\textbf{B}_u^{ - 1}}$ is a positive semidefinite matrix, ${\textbf{x}_{a(t)}}^T{\textbf{B}_u^{ - 1}}{\textbf{x}_{a(t)}}$ is non-negative. It is easy to see the change of the expectation of new reward is decided by de-biased user's feedback $r_a'$. For example, if a user gives a positive feedback on an arm $a(t)$, next time the algorithm tends to score higher for $a(t)$ as well as the arms with similar embeddings with $a(t)$. And if the feedback is negative the score will be relatively smaller. This strategy helps the algorithm to do exploitation by remembering past experience and taking actions accordingly. As for the second part ${\textbf{B}_u^{ - 1}}$ in the update formula, we take it as a regularization term to confine the updating of $\bm{\mu}_u$ (cause the decrease of diagonal elements of ${\textbf{B}_u^{ - 1}}$).

We also need to mention that since we regard all attributes and items as equivalent arms, we can estimate the expected rewards of asking attribute or recommending items by updating the parameters in the same way, thus enable the algorithm to choose the action intelligently according to different situations. The model will learn to identify the benefit of taking different actions and make the best choice accordingly.

\bibliographystyle{ACM-Reference-Format}
\bibliography{reference.bbl}


\begin{thebibliography}{55}


\ifx \showCODEN    \undefined \def \showCODEN     #1{\unskip}     \fi
\ifx \showDOI      \undefined \def \showDOI       #1{#1}\fi
\ifx \showISBNx    \undefined \def \showISBNx     #1{\unskip}     \fi
\ifx \showISBNxiii \undefined \def \showISBNxiii  #1{\unskip}     \fi
\ifx \showISSN     \undefined \def \showISSN      #1{\unskip}     \fi
\ifx \showLCCN     \undefined \def \showLCCN      #1{\unskip}     \fi
\ifx \shownote     \undefined \def \shownote      #1{#1}          \fi
\ifx \showarticletitle \undefined \def \showarticletitle #1{#1}   \fi
\ifx \showURL      \undefined \def \showURL       {\relax}        \fi
\providecommand\bibfield[2]{#2}
\providecommand\bibinfo[2]{#2}
\providecommand\natexlab[1]{#1}
\providecommand\showeprint[2][]{arXiv:#2}

\bibitem[\protect\citeauthoryear{Agrawal and Goyal}{Agrawal and Goyal}{2013}]%
        {agrawal2013thompson}
\bibfield{author}{\bibinfo{person}{Shipra Agrawal} {and} \bibinfo{person}{Navin
  Goyal}.} \bibinfo{year}{2013}\natexlab{}.
\newblock \showarticletitle{Thompson sampling for contextual bandits with
  linear payoffs}. In \bibinfo{booktitle}{\emph{ICML}}.
  \bibinfo{pages}{127--135}.
\newblock


\bibitem[\protect\citeauthoryear{Aliannejadi, Zamani, Crestani, and
  Croft}{Aliannejadi et~al\mbox{.}}{2019}]%
        {AskingClarifying}
\bibfield{author}{\bibinfo{person}{Mohammad Aliannejadi},
  \bibinfo{person}{Hamed Zamani}, \bibinfo{person}{Fabio Crestani}, {and}
  \bibinfo{person}{W.~Bruce Croft}.} \bibinfo{year}{2019}\natexlab{}.
\newblock \showarticletitle{Asking Clarifying Questions in Open-Domain
  Information-Seeking Conversations}. In \bibinfo{booktitle}{\emph{SIGIR}}.
  \bibinfo{pages}{475--484}.
\newblock


\bibitem[\protect\citeauthoryear{Auer, Cesa-Bianchi, and Fischer}{Auer
  et~al\mbox{.}}{2002a}]%
        {auer2002finite}
\bibfield{author}{\bibinfo{person}{Peter Auer}, \bibinfo{person}{Nicolo
  Cesa-Bianchi}, {and} \bibinfo{person}{Paul Fischer}.}
  \bibinfo{year}{2002}\natexlab{a}.
\newblock \showarticletitle{Finite-time analysis of the multiarmed bandit
  problem}.
\newblock \bibinfo{journal}{\emph{Machine learning}} \bibinfo{volume}{47},
  \bibinfo{number}{2-3} (\bibinfo{year}{2002}), \bibinfo{pages}{235--256}.
\newblock


\bibitem[\protect\citeauthoryear{Auer, Cesa-Bianchi, and Fischer}{Auer
  et~al\mbox{.}}{2002b}]%
        {UCB1}
\bibfield{author}{\bibinfo{person}{Peter Auer}, \bibinfo{person}{Nicol\`{o}
  Cesa-Bianchi}, {and} \bibinfo{person}{Paul Fischer}.}
  \bibinfo{year}{2002}\natexlab{b}.
\newblock \showarticletitle{Finite-time Analysis of the Multiarmed Bandit
  Problem}.
\newblock \bibinfo{journal}{\emph{Mach. Learn.}} \bibinfo{volume}{47},
  \bibinfo{number}{2-3} (\bibinfo{date}{May} \bibinfo{year}{2002}),
  \bibinfo{pages}{235--256}.
\newblock
\showISSN{0885-6125}
\urldef\tempurl%
\url{https://doi.org/10.1023/A:1013689704352}
\showDOI{\tempurl}


\bibitem[\protect\citeauthoryear{Averjanova, Ricci, and Nguyen}{Averjanova
  et~al\mbox{.}}{2008}]%
        {averjanova2008map}
\bibfield{author}{\bibinfo{person}{Olga Averjanova}, \bibinfo{person}{Francesco
  Ricci}, {and} \bibinfo{person}{Quang~Nhat Nguyen}.}
  \bibinfo{year}{2008}\natexlab{}.
\newblock \showarticletitle{Map-based interaction with a conversational mobile
  recommender system}. In \bibinfo{booktitle}{\emph{2008 The Second
  International Conference on Mobile Ubiquitous Computing, Systems, Services
  and Technologies}}. IEEE, \bibinfo{pages}{212--218}.
\newblock


\bibitem[\protect\citeauthoryear{Chapelle and Li}{Chapelle and Li}{2011}]%
        {chapelle2011empirical}
\bibfield{author}{\bibinfo{person}{Olivier Chapelle} {and}
  \bibinfo{person}{Lihong Li}.} \bibinfo{year}{2011}\natexlab{}.
\newblock \showarticletitle{An empirical evaluation of thompson sampling}. In
  \bibinfo{booktitle}{\emph{NIPS}}. \bibinfo{pages}{2249--2257}.
\newblock


\bibitem[\protect\citeauthoryear{Chen, Zhang, He, Nie, Liu, and Chua}{Chen
  et~al\mbox{.}}{2017}]%
        {ACF}
\bibfield{author}{\bibinfo{person}{Jingyuan Chen}, \bibinfo{person}{Hanwang
  Zhang}, \bibinfo{person}{Xiangnan He}, \bibinfo{person}{Liqiang Nie},
  \bibinfo{person}{Wei Liu}, {and} \bibinfo{person}{Tat{-}Seng Chua}.}
  \bibinfo{year}{2017}\natexlab{}.
\newblock \showarticletitle{Attentive Collaborative Filtering: Multimedia
  Recommendation with Item- and Component-Level Attention}. In
  \bibinfo{booktitle}{\emph{SIGIR}}. \bibinfo{pages}{335--344}.
\newblock


\bibitem[\protect\citeauthoryear{Chen, Lin, Zhang, Ding, Cen, Yang, and
  Tang}{Chen et~al\mbox{.}}{2019}]%
        {chen-etal-2019-towards}
\bibfield{author}{\bibinfo{person}{Qibin Chen}, \bibinfo{person}{Junyang Lin},
  \bibinfo{person}{Yichang Zhang}, \bibinfo{person}{Ming Ding},
  \bibinfo{person}{Yukuo Cen}, \bibinfo{person}{Hongxia Yang}, {and}
  \bibinfo{person}{Jie Tang}.} \bibinfo{year}{2019}\natexlab{}.
\newblock \showarticletitle{Towards Knowledge-Based Recommender Dialog System}.
  In \bibinfo{booktitle}{\emph{EMNLP-IJCNLP}}. \bibinfo{pages}{1803--1813}.
\newblock


\bibitem[\protect\citeauthoryear{Cheng, Koc, Harmsen, Shaked, Chandra, Aradhye,
  Anderson, Corrado, Chai, Ispir, et~al\mbox{.}}{Cheng et~al\mbox{.}}{2016}]%
        {cheng2016wide}
\bibfield{author}{\bibinfo{person}{Heng-Tze Cheng}, \bibinfo{person}{Levent
  Koc}, \bibinfo{person}{Jeremiah Harmsen}, \bibinfo{person}{Tal Shaked},
  \bibinfo{person}{Tushar Chandra}, \bibinfo{person}{Hrishi Aradhye},
  \bibinfo{person}{Glen Anderson}, \bibinfo{person}{Greg Corrado},
  \bibinfo{person}{Wei Chai}, \bibinfo{person}{Mustafa Ispir}, {et~al\mbox{.}}}
  \bibinfo{year}{2016}\natexlab{}.
\newblock \showarticletitle{Wide \& deep learning for recommender systems}. In
  \bibinfo{booktitle}{\emph{DLRS@RecSys}}. \bibinfo{pages}{7--10}.
\newblock


\bibitem[\protect\citeauthoryear{Christakopoulou, Beutel, Li, Jain, and
  Chi}{Christakopoulou et~al\mbox{.}}{2018}]%
        {christakopoulou2018q}
\bibfield{author}{\bibinfo{person}{Konstantina Christakopoulou},
  \bibinfo{person}{Alex Beutel}, \bibinfo{person}{Rui Li},
  \bibinfo{person}{Sagar Jain}, {and} \bibinfo{person}{Ed~H Chi}.}
  \bibinfo{year}{2018}\natexlab{}.
\newblock \showarticletitle{Q\&R: A Two-Stage Approach toward Interactive
  Recommendation}. In \bibinfo{booktitle}{\emph{SIGKDD}}.
  \bibinfo{pages}{139--148}.
\newblock


\bibitem[\protect\citeauthoryear{Christakopoulou, Radlinski, and
  Hofmann}{Christakopoulou et~al\mbox{.}}{2016}]%
        {christakopoulou2016towards}
\bibfield{author}{\bibinfo{person}{Konstantina Christakopoulou},
  \bibinfo{person}{Filip Radlinski}, {and} \bibinfo{person}{Katja Hofmann}.}
  \bibinfo{year}{2016}\natexlab{}.
\newblock \showarticletitle{Towards conversational recommender systems}. In
  \bibinfo{booktitle}{\emph{KDD}}. \bibinfo{pages}{815--824}.
\newblock


\bibitem[\protect\citeauthoryear{Dietz, Myftija, and W{\"o}rndl}{Dietz
  et~al\mbox{.}}{2019}]%
        {dietz2019designing}
\bibfield{author}{\bibinfo{person}{Linus~W Dietz}, \bibinfo{person}{Saadi
  Myftija}, {and} \bibinfo{person}{Wolfgang W{\"o}rndl}.}
  \bibinfo{year}{2019}\natexlab{}.
\newblock \showarticletitle{Designing a conversational travel recommender
  system based on data-driven destination characterization}. In
  \bibinfo{booktitle}{\emph{ACM RecSys workshop on recommenders in tourism}}.
  \bibinfo{pages}{17--21}.
\newblock


\bibitem[\protect\citeauthoryear{Gentile, Li, Kar, Karatzoglou, Etrue, and
  Zappella}{Gentile et~al\mbox{.}}{2016}]%
        {gentile2016context}
\bibfield{author}{\bibinfo{person}{Claudio Gentile}, \bibinfo{person}{Shuai
  Li}, \bibinfo{person}{Purushottam Kar}, \bibinfo{person}{Alexandros
  Karatzoglou}, \bibinfo{person}{Evans Etrue}, {and} \bibinfo{person}{Giovanni
  Zappella}.} \bibinfo{year}{2016}\natexlab{}.
\newblock \showarticletitle{On context-dependent clustering of bandits}. In
  \bibinfo{booktitle}{\emph{ICML}}. \bibinfo{pages}{1253--1262}.
\newblock


\bibitem[\protect\citeauthoryear{Graepel, Candela, Borchert, and
  Herbrich}{Graepel et~al\mbox{.}}{2010}]%
        {graepel2010web}
\bibfield{author}{\bibinfo{person}{Thore Graepel},
  \bibinfo{person}{Joaquin~Quinonero Candela}, \bibinfo{person}{Thomas
  Borchert}, {and} \bibinfo{person}{Ralf Herbrich}.}
  \bibinfo{year}{2010}\natexlab{}.
\newblock \showarticletitle{Web-scale bayesian click-through rate prediction
  for sponsored search advertising in microsoft's bing search engine}. In
  \bibinfo{booktitle}{\emph{ICML}}.
\newblock


\bibitem[\protect\citeauthoryear{Granmo}{Granmo}{2010}]%
        {granmo2010solving}
\bibfield{author}{\bibinfo{person}{Ole-Christoffer Granmo}.}
  \bibinfo{year}{2010}\natexlab{}.
\newblock \showarticletitle{Solving two-armed Bernoulli bandit problems using a
  Bayesian learning automaton}.
\newblock \bibinfo{journal}{\emph{International Journal of Intelligent
  Computing and Cybernetics}} \bibinfo{volume}{3}, \bibinfo{number}{2}
  (\bibinfo{year}{2010}), \bibinfo{pages}{207--234}.
\newblock


\bibitem[\protect\citeauthoryear{Guo, Tang, Ye, Li, and He}{Guo
  et~al\mbox{.}}{2017}]%
        {guo2017deepfm}
\bibfield{author}{\bibinfo{person}{Huifeng Guo}, \bibinfo{person}{Ruiming
  Tang}, \bibinfo{person}{Yunming Ye}, \bibinfo{person}{Zhenguo Li}, {and}
  \bibinfo{person}{Xiuqiang He}.} \bibinfo{year}{2017}\natexlab{}.
\newblock \showarticletitle{DeepFM: a factorization-machine based neural
  network for CTR prediction}.
\newblock \bibinfo{journal}{\emph{IJCAI}} (\bibinfo{year}{2017}).
\newblock


\bibitem[\protect\citeauthoryear{He and Chua}{He and Chua}{2017}]%
        {he2017neural}
\bibfield{author}{\bibinfo{person}{Xiangnan He} {and} \bibinfo{person}{Tat-Seng
  Chua}.} \bibinfo{year}{2017}\natexlab{}.
\newblock \showarticletitle{Neural factorization machines for sparse predictive
  analytics}. In \bibinfo{booktitle}{\emph{SIGIR}}. \bibinfo{pages}{355--364}.
\newblock


\bibitem[\protect\citeauthoryear{He, Deng, Wang, Li, Zhang, and Wang}{He
  et~al\mbox{.}}{2020}]%
        {LightGCN}
\bibfield{author}{\bibinfo{person}{Xiangnan He}, \bibinfo{person}{Kuan Deng},
  \bibinfo{person}{Xiang Wang}, \bibinfo{person}{Yan Li},
  \bibinfo{person}{Yongdong Zhang}, {and} \bibinfo{person}{Meng Wang}.}
  \bibinfo{year}{2020}\natexlab{}.
\newblock \showarticletitle{LightGCN: Simplifying and Powering Graph
  Convolution Network for Recommendation}. In
  \bibinfo{booktitle}{\emph{SIGIR}}.
\newblock


\bibitem[\protect\citeauthoryear{He, Liao, Zhang, Nie, Hu, and Chua}{He
  et~al\mbox{.}}{2017}]%
        {NCF}
\bibfield{author}{\bibinfo{person}{Xiangnan He}, \bibinfo{person}{Lizi Liao},
  \bibinfo{person}{Hanwang Zhang}, \bibinfo{person}{Liqiang Nie},
  \bibinfo{person}{Xia Hu}, {and} \bibinfo{person}{Tat{-}Seng Chua}.}
  \bibinfo{year}{2017}\natexlab{}.
\newblock \showarticletitle{Neural Collaborative Filtering}. In
  \bibinfo{booktitle}{\emph{WWW}}. \bibinfo{pages}{173--182}.
\newblock


\bibitem[\protect\citeauthoryear{He, Zhang, Kan, and Chua}{He
  et~al\mbox{.}}{2016}]%
        {fastMF}
\bibfield{author}{\bibinfo{person}{Xiangnan He}, \bibinfo{person}{Hanwang
  Zhang}, \bibinfo{person}{Min-Yen Kan}, {and} \bibinfo{person}{Tat-Seng
  Chua}.} \bibinfo{year}{2016}\natexlab{}.
\newblock \showarticletitle{Fast matrix factorization for online recommendation
  with implicit feedback}. In \bibinfo{booktitle}{\emph{SIGIR}}.
  \bibinfo{pages}{549--558}.
\newblock


\bibitem[\protect\citeauthoryear{Hong, Huang, Chin, Yen, and Lin}{Hong
  et~al\mbox{.}}{2010}]%
        {hong2010interactive}
\bibfield{author}{\bibinfo{person}{Zeng-Wei Hong}, \bibinfo{person}{Rui-Tang
  Huang}, \bibinfo{person}{Kai-Yi Chin}, \bibinfo{person}{Chia-Chi Yen}, {and}
  \bibinfo{person}{Jim-Min Lin}.} \bibinfo{year}{2010}\natexlab{}.
\newblock \showarticletitle{An interactive agent system for supporting
  knowledge-based recommendation: a case study on an e-novel recommender
  system}. In \bibinfo{booktitle}{\emph{Proceedings of the 4th International
  Conference on Uniquitous Information Management and Communication}}.
  \bibinfo{pages}{1--8}.
\newblock


\bibitem[\protect\citeauthoryear{Jannach, Manzoor, Cai, and Chen}{Jannach
  et~al\mbox{.}}{2020}]%
        {abs-2004-00646}
\bibfield{author}{\bibinfo{person}{Dietmar Jannach}, \bibinfo{person}{Ahtsham
  Manzoor}, \bibinfo{person}{Wanling Cai}, {and} \bibinfo{person}{Li Chen}.}
  \bibinfo{year}{2020}\natexlab{}.
\newblock \showarticletitle{A Survey on Conversational Recommender Systems}.
\newblock \bibinfo{journal}{\emph{CoRR}} (\bibinfo{year}{2020}).
\newblock


\bibitem[\protect\citeauthoryear{Jiang, Ren, Monz, and de~Rijke}{Jiang
  et~al\mbox{.}}{2019}]%
        {jiang-2019-improving}
\bibfield{author}{\bibinfo{person}{Shaojie Jiang}, \bibinfo{person}{Pengjie
  Ren}, \bibinfo{person}{Christof Monz}, {and} \bibinfo{person}{Maarten de
  Rijke}.} \bibinfo{year}{2019}\natexlab{}.
\newblock \showarticletitle{Improving Neural Response Diversity with
  Frequency-Aware Cross-Entropy Loss}. In \bibinfo{booktitle}{\emph{The Web
  Conference 2019}}. \bibinfo{publisher}{ACM}, \bibinfo{pages}{2879--2885}.
\newblock


\bibitem[\protect\citeauthoryear{Jin, Lei, Ren, Chen, Liang, Zhao, and Yin}{Jin
  et~al\mbox{.}}{2018}]%
        {jin2018explicit}
\bibfield{author}{\bibinfo{person}{Xisen Jin}, \bibinfo{person}{Wenqiang Lei},
  \bibinfo{person}{Zhaochun Ren}, \bibinfo{person}{Hongshen Chen},
  \bibinfo{person}{Shangsong Liang}, \bibinfo{person}{Yihong Zhao}, {and}
  \bibinfo{person}{Dawei Yin}.} \bibinfo{year}{2018}\natexlab{}.
\newblock \showarticletitle{Explicit State Tracking with Semi-Supervisionfor
  Neural Dialogue Generation}. In \bibinfo{booktitle}{\emph{CIKM}}. ACM,
  \bibinfo{pages}{1403--1412}.
\newblock


\bibitem[\protect\citeauthoryear{Koren, Bell, and Volinsky}{Koren
  et~al\mbox{.}}{2009}]%
        {koren2009matrix}
\bibfield{author}{\bibinfo{person}{Yehuda Koren}, \bibinfo{person}{Robert~M.
  Bell}, {and} \bibinfo{person}{Chris Volinsky}.}
  \bibinfo{year}{2009}\natexlab{}.
\newblock \showarticletitle{Matrix Factorization Techniques for Recommender
  Systems}.
\newblock \bibinfo{journal}{\emph{{IEEE} Computer}} (\bibinfo{year}{2009}),
  \bibinfo{pages}{30--37}.
\newblock


\bibitem[\protect\citeauthoryear{Lai and Robbins}{Lai and Robbins}{1985}]%
        {lai1985asymptotically}
\bibfield{author}{\bibinfo{person}{Tze~Leung Lai} {and}
  \bibinfo{person}{Herbert Robbins}.} \bibinfo{year}{1985}\natexlab{}.
\newblock \showarticletitle{Asymptotically efficient adaptive allocation
  rules}.
\newblock \bibinfo{journal}{\emph{Advances in applied mathematics}}
  \bibinfo{volume}{6}, \bibinfo{number}{1} (\bibinfo{year}{1985}),
  \bibinfo{pages}{4--22}.
\newblock


\bibitem[\protect\citeauthoryear{Lei, He, Miao, Wu, Hong, Kan, and Chua}{Lei
  et~al\mbox{.}}{2020}]%
        {lei20estimation}
\bibfield{author}{\bibinfo{person}{Wenqiang Lei}, \bibinfo{person}{Xiangnan
  He}, \bibinfo{person}{Yisong Miao}, \bibinfo{person}{Qingyun Wu},
  \bibinfo{person}{Richang Hong}, \bibinfo{person}{Min-Yen Kan}, {and}
  \bibinfo{person}{Tat-Seng Chua}.} \bibinfo{year}{2020}\natexlab{}.
\newblock \showarticletitle{Estimation--Action--Reflection: Towards Deep
  Interaction Between Conversational and Recommender Systems}. In
  \bibinfo{booktitle}{\emph{WSDM}}.
\newblock


\bibitem[\protect\citeauthoryear{Lei, Jin, Kan, Ren, He, and Yin}{Lei
  et~al\mbox{.}}{2018}]%
        {acl18/sequicity}
\bibfield{author}{\bibinfo{person}{Wenqiang Lei}, \bibinfo{person}{Xisen Jin},
  \bibinfo{person}{Min{-}Yen Kan}, \bibinfo{person}{Zhaochun Ren},
  \bibinfo{person}{Xiangnan He}, {and} \bibinfo{person}{Dawei Yin}.}
  \bibinfo{year}{2018}\natexlab{}.
\newblock \showarticletitle{Sequicity: Simplifying Task-oriented Dialogue
  Systems with Single Sequence-to-Sequence Architectures}. In
  \bibinfo{booktitle}{\emph{ACL}}. \bibinfo{pages}{1437--1447}.
\newblock


\bibitem[\protect\citeauthoryear{Li, Chu, Langford, and Schapire}{Li
  et~al\mbox{.}}{2010}]%
        {li2010contextual}
\bibfield{author}{\bibinfo{person}{Lihong Li}, \bibinfo{person}{Wei Chu},
  \bibinfo{person}{John Langford}, {and} \bibinfo{person}{Robert~E Schapire}.}
  \bibinfo{year}{2010}\natexlab{}.
\newblock \showarticletitle{A contextual-bandit approach to personalized news
  article recommendation}. In \bibinfo{booktitle}{\emph{WWW}}.
  \bibinfo{pages}{661--670}.
\newblock


\bibitem[\protect\citeauthoryear{Li and Karahanna}{Li and Karahanna}{2015}]%
        {li2015online}
\bibfield{author}{\bibinfo{person}{Seth~Siyuan Li} {and} \bibinfo{person}{Elena
  Karahanna}.} \bibinfo{year}{2015}\natexlab{}.
\newblock \showarticletitle{Online recommendation systems in a B2C E-commerce
  context: a review and future directions}.
\newblock \bibinfo{journal}{\emph{Journal of the Association for Information
  Systems}} \bibinfo{volume}{16}, \bibinfo{number}{2} (\bibinfo{year}{2015}),
  \bibinfo{pages}{72}.
\newblock


\bibitem[\protect\citeauthoryear{May and Leslie}{May and Leslie}{2011}]%
        {may2011simulation}
\bibfield{author}{\bibinfo{person}{Benedict~C May} {and}
  \bibinfo{person}{David~S Leslie}.} \bibinfo{year}{2011}\natexlab{}.
\newblock \showarticletitle{Simulation studies in optimistic Bayesian sampling
  in contextual-bandit problems}.
\newblock \bibinfo{journal}{\emph{Statistics Group, Department of Mathematics,
  University of Bristol}} \bibinfo{volume}{11}, \bibinfo{number}{02}
  (\bibinfo{year}{2011}).
\newblock


\bibitem[\protect\citeauthoryear{McCarthy, Reilly, McGinty, and Smyth}{McCarthy
  et~al\mbox{.}}{2004}]%
        {mccarthy2004dynamic}
\bibfield{author}{\bibinfo{person}{Kevin McCarthy}, \bibinfo{person}{James
  Reilly}, \bibinfo{person}{Lorraine McGinty}, {and} \bibinfo{person}{Barry
  Smyth}.} \bibinfo{year}{2004}\natexlab{}.
\newblock \showarticletitle{On the dynamic generation of compound critiques in
  conversational recommender systems}. In
  \bibinfo{booktitle}{\emph{International Conference on Adaptive Hypermedia and
  Adaptive Web-Based Systems}}. Springer, \bibinfo{pages}{176--184}.
\newblock


\bibitem[\protect\citeauthoryear{Murphy}{Murphy}{2012}]%
        {murphy2012machine}
\bibfield{author}{\bibinfo{person}{Kevin~P Murphy}.}
  \bibinfo{year}{2012}\natexlab{}.
\newblock \bibinfo{booktitle}{\emph{Machine learning: a probabilistic
  perspective}}.
\newblock \bibinfo{publisher}{MIT press}.
\newblock
\showISBNx{0262018020}


\bibitem[\protect\citeauthoryear{Osband and Van~Roy}{Osband and
  Van~Roy}{2017}]%
        {osband2017optimistic}
\bibfield{author}{\bibinfo{person}{Ian Osband} {and} \bibinfo{person}{Benjamin
  Van~Roy}.} \bibinfo{year}{2017}\natexlab{}.
\newblock \showarticletitle{On optimistic versus randomized exploration in
  reinforcement learning}.
\newblock \bibinfo{journal}{\emph{arXiv preprint arXiv:1706.04241}}
  (\bibinfo{year}{2017}).
\newblock


\bibitem[\protect\citeauthoryear{Priyogi}{Priyogi}{2019}]%
        {priyogi2019preference}
\bibfield{author}{\bibinfo{person}{Bilih Priyogi}.}
  \bibinfo{year}{2019}\natexlab{}.
\newblock \showarticletitle{Preference Elicitation Strategy for Conversational
  Recommender System}. In \bibinfo{booktitle}{\emph{WSDM}}. ACM,
  \bibinfo{pages}{824--825}.
\newblock


\bibitem[\protect\citeauthoryear{Radlinski, Balog, Byrne, and
  Krishnamoorthi}{Radlinski et~al\mbox{.}}{2019}]%
        {radlinski-etal-2019-coached}
\bibfield{author}{\bibinfo{person}{Filip Radlinski}, \bibinfo{person}{Krisztian
  Balog}, \bibinfo{person}{Bill Byrne}, {and} \bibinfo{person}{Karthik
  Krishnamoorthi}.} \bibinfo{year}{2019}\natexlab{}.
\newblock \showarticletitle{Coached Conversational Preference Elicitation: A
  Case Study in Understanding Movie Preferences}. In
  \bibinfo{booktitle}{\emph{Proceedings of the 20th Annual SIGdial Meeting on
  Discourse and Dialogue}}. \bibinfo{pages}{353--360}.
\newblock


\bibitem[\protect\citeauthoryear{Ren, Chen, Monz, Ma, and de~Rijke}{Ren
  et~al\mbox{.}}{2020}]%
        {ren-2020-thinking}
\bibfield{author}{\bibinfo{person}{Pengjie Ren}, \bibinfo{person}{Zhumin Chen},
  \bibinfo{person}{Christof Monz}, \bibinfo{person}{Jun Ma}, {and}
  \bibinfo{person}{Maarten de Rijke}.} \bibinfo{year}{2020}\natexlab{}.
\newblock \showarticletitle{Thinking Globally, Acting Locally: Distantly
  Supervised Global-to-Local Knowledge Selection for Background Based
  Conversation}. In \bibinfo{booktitle}{\emph{Thirty-Fourth AAAI Conference on
  Artificial Intelligence (AAAI-20)}}. \bibinfo{publisher}{AAAI}.
\newblock


\bibitem[\protect\citeauthoryear{Rendle}{Rendle}{2010}]%
        {FM}
\bibfield{author}{\bibinfo{person}{Steffen Rendle}.}
  \bibinfo{year}{2010}\natexlab{}.
\newblock \showarticletitle{Factorization machines}. In
  \bibinfo{booktitle}{\emph{ICDM}}. \bibinfo{pages}{995--1000}.
\newblock


\bibitem[\protect\citeauthoryear{Rendle, Freudenthaler, Gantner, and
  Schmidt-Thieme}{Rendle et~al\mbox{.}}{2009}]%
        {BPR}
\bibfield{author}{\bibinfo{person}{Steffen Rendle}, \bibinfo{person}{Christoph
  Freudenthaler}, \bibinfo{person}{Zeno Gantner}, {and} \bibinfo{person}{Lars
  Schmidt-Thieme}.} \bibinfo{year}{2009}\natexlab{}.
\newblock \showarticletitle{BPR: Bayesian personalized ranking from implicit
  feedback}. In \bibinfo{booktitle}{\emph{UAI}}.
\newblock


\bibitem[\protect\citeauthoryear{Russo and Van~Roy}{Russo and Van~Roy}{2014}]%
        {russo2014learning}
\bibfield{author}{\bibinfo{person}{Daniel Russo} {and}
  \bibinfo{person}{Benjamin Van~Roy}.} \bibinfo{year}{2014}\natexlab{}.
\newblock \showarticletitle{Learning to optimize via posterior sampling}.
\newblock \bibinfo{journal}{\emph{Mathematics of Operations Research}}
  (\bibinfo{year}{2014}), \bibinfo{pages}{1221--1243}.
\newblock


\bibitem[\protect\citeauthoryear{Russo, Van~Roy, Kazerouni, Osband, Wen,
  et~al\mbox{.}}{Russo et~al\mbox{.}}{2018}]%
        {russo2018tutorial}
\bibfield{author}{\bibinfo{person}{Daniel~J Russo}, \bibinfo{person}{Benjamin
  Van~Roy}, \bibinfo{person}{Abbas Kazerouni}, \bibinfo{person}{Ian Osband},
  \bibinfo{person}{Zheng Wen}, {et~al\mbox{.}}}
  \bibinfo{year}{2018}\natexlab{}.
\newblock \showarticletitle{A tutorial on thompson sampling}.
\newblock \bibinfo{journal}{\emph{Foundations and Trends{\textregistered} in
  Machine Learning}} (\bibinfo{year}{2018}), \bibinfo{pages}{1--96}.
\newblock


\bibitem[\protect\citeauthoryear{Sardella, Biancalana, Micarelli, and
  Sansonetti}{Sardella et~al\mbox{.}}{2019}]%
        {sardella2019approach}
\bibfield{author}{\bibinfo{person}{Nicola Sardella}, \bibinfo{person}{Claudio
  Biancalana}, \bibinfo{person}{Alessandro Micarelli}, {and}
  \bibinfo{person}{Giuseppe Sansonetti}.} \bibinfo{year}{2019}\natexlab{}.
\newblock \showarticletitle{An Approach to Conversational Recommendation of
  Restaurants}. In \bibinfo{booktitle}{\emph{ICHCI}}. Springer,
  \bibinfo{pages}{123--130}.
\newblock


\bibitem[\protect\citeauthoryear{Shimazu}{Shimazu}{2002}]%
        {shimazu2002expertclerk}
\bibfield{author}{\bibinfo{person}{Hideo Shimazu}.}
  \bibinfo{year}{2002}\natexlab{}.
\newblock \showarticletitle{ExpertClerk: A Conversational Case-Based Reasoning
  Tool forDeveloping Salesclerk Agents in E-Commerce Webshops}.
\newblock \bibinfo{journal}{\emph{Artificial Intelligence Review}}
  \bibinfo{volume}{18}, \bibinfo{number}{3-4} (\bibinfo{year}{2002}),
  \bibinfo{pages}{223--244}.
\newblock


\bibitem[\protect\citeauthoryear{Sun and Zhang}{Sun and Zhang}{2018}]%
        {sun2018conversational}
\bibfield{author}{\bibinfo{person}{Yueming Sun} {and} \bibinfo{person}{Yi
  Zhang}.} \bibinfo{year}{2018}\natexlab{}.
\newblock \showarticletitle{Conversational recommender system}. In
  \bibinfo{booktitle}{\emph{SIGIR}}. \bibinfo{pages}{235--244}.
\newblock


\bibitem[\protect\citeauthoryear{Thompson, Goker, and Langley}{Thompson
  et~al\mbox{.}}{2004}]%
        {thompson2004personalized}
\bibfield{author}{\bibinfo{person}{Cynthia~A Thompson},
  \bibinfo{person}{Mehmet~H Goker}, {and} \bibinfo{person}{Pat Langley}.}
  \bibinfo{year}{2004}\natexlab{}.
\newblock \showarticletitle{A personalized system for conversational
  recommendations}.
\newblock \bibinfo{journal}{\emph{Journal of Artificial Intelligence Research}}
   \bibinfo{volume}{21} (\bibinfo{year}{2004}), \bibinfo{pages}{393--428}.
\newblock


\bibitem[\protect\citeauthoryear{van~den Berg, Kipf, and Welling}{van~den Berg
  et~al\mbox{.}}{2017}]%
        {vdberg2017graph}
\bibfield{author}{\bibinfo{person}{Rianne van~den Berg},
  \bibinfo{person}{Thomas~N Kipf}, {and} \bibinfo{person}{Max Welling}.}
  \bibinfo{year}{2017}\natexlab{}.
\newblock \showarticletitle{Graph Convolutional Matrix Completion}. In
  \bibinfo{booktitle}{\emph{KDD}}.
\newblock


\bibitem[\protect\citeauthoryear{Wang, Wu, and Wang}{Wang
  et~al\mbox{.}}{2017}]%
        {wang2017factorization}
\bibfield{author}{\bibinfo{person}{Huazheng Wang}, \bibinfo{person}{Qingyun
  Wu}, {and} \bibinfo{person}{Hongning Wang}.} \bibinfo{year}{2017}\natexlab{}.
\newblock \showarticletitle{Factorization Bandits for Interactive
  Recommendation.}. In \bibinfo{booktitle}{\emph{AAAI}}.
  \bibinfo{pages}{2695--2702}.
\newblock


\bibitem[\protect\citeauthoryear{Wang, He, Wang, Feng, and Chua}{Wang
  et~al\mbox{.}}{2019}]%
        {wang2019neural}
\bibfield{author}{\bibinfo{person}{Xiang Wang}, \bibinfo{person}{Xiangnan He},
  \bibinfo{person}{Meng Wang}, \bibinfo{person}{Fuli Feng}, {and}
  \bibinfo{person}{Tat-Seng Chua}.} \bibinfo{year}{2019}\natexlab{}.
\newblock \showarticletitle{Neural graph collaborative filtering}. In
  \bibinfo{booktitle}{\emph{SIGIR}}. \bibinfo{pages}{165--174}.
\newblock


\bibitem[\protect\citeauthoryear{Wei, Liu, Zheng, Zhang, Wang, and Wu}{Wei
  et~al\mbox{.}}{2015}]%
        {wei2015df}
\bibfield{author}{\bibinfo{person}{Bifan Wei}, \bibinfo{person}{Jun Liu},
  \bibinfo{person}{Qinghua Zheng}, \bibinfo{person}{Wei Zhang},
  \bibinfo{person}{Chenchen Wang}, {and} \bibinfo{person}{Bei Wu}.}
  \bibinfo{year}{2015}\natexlab{}.
\newblock \showarticletitle{DF-Miner: Domain-specific facet mining by
  leveraging the hyperlink structure of Wikipedia}.
\newblock \bibinfo{journal}{\emph{Knowledge-Based Systems}}
  \bibinfo{volume}{77} (\bibinfo{year}{2015}), \bibinfo{pages}{80--91}.
\newblock


\bibitem[\protect\citeauthoryear{Wu, Iyer, and Wang}{Wu et~al\mbox{.}}{2018}]%
        {wu2018learning}
\bibfield{author}{\bibinfo{person}{Qingyun Wu}, \bibinfo{person}{Naveen Iyer},
  {and} \bibinfo{person}{Hongning Wang}.} \bibinfo{year}{2018}\natexlab{}.
\newblock \showarticletitle{Learning Contextual Bandits in a Non-stationary
  Environment}. In \bibinfo{booktitle}{\emph{SIGIR}}.
  \bibinfo{pages}{495--504}.
\newblock


\bibitem[\protect\citeauthoryear{Wu, Wang, Gu, and Wang}{Wu
  et~al\mbox{.}}{2016}]%
        {wu2016contextual}
\bibfield{author}{\bibinfo{person}{Qingyun Wu}, \bibinfo{person}{Huazheng
  Wang}, \bibinfo{person}{Quanquan Gu}, {and} \bibinfo{person}{Hongning Wang}.}
  \bibinfo{year}{2016}\natexlab{}.
\newblock \showarticletitle{Contextual bandits in a collaborative environment}.
  In \bibinfo{booktitle}{\emph{SIGIR}}. \bibinfo{pages}{529--538}.
\newblock


\bibitem[\protect\citeauthoryear{Yu, Shen, and Jin}{Yu et~al\mbox{.}}{2019}]%
        {yu2019visual}
\bibfield{author}{\bibinfo{person}{Tong Yu}, \bibinfo{person}{Yilin Shen},
  {and} \bibinfo{person}{Hongxia Jin}.} \bibinfo{year}{2019}\natexlab{}.
\newblock \showarticletitle{An Visual Dialog Augmented Interactive Recommender
  System}. In \bibinfo{booktitle}{\emph{SIGKDD}}. ACM,
  \bibinfo{pages}{157--165}.
\newblock


\bibitem[\protect\citeauthoryear{Zamani, Dumais, Craswell, Bennett, and
  Lueck}{Zamani et~al\mbox{.}}{2020}]%
        {GeneratingClarifying}
\bibfield{author}{\bibinfo{person}{Hamed Zamani}, \bibinfo{person}{Susan~T.
  Dumais}, \bibinfo{person}{Nick Craswell}, \bibinfo{person}{Paul~N. Bennett},
  {and} \bibinfo{person}{Gord Lueck}.} \bibinfo{year}{2020}\natexlab{}.
\newblock \showarticletitle{Generating Clarifying Questions for Information
  Retrieval}. In \bibinfo{booktitle}{\emph{WWW}}.
\newblock


\bibitem[\protect\citeauthoryear{Zhang, Xie, Li, and Lui}{Zhang
  et~al\mbox{.}}{2020}]%
        {zhang2020toward}
\bibfield{author}{\bibinfo{person}{Xiaoying Zhang}, \bibinfo{person}{Hong Xie},
  \bibinfo{person}{Hang Li}, {and} \bibinfo{person}{John Lui}.}
  \bibinfo{year}{2020}\natexlab{}.
\newblock \showarticletitle{Conversational Contextual Bandit: Algorithm and
  Application}. In \bibinfo{booktitle}{\emph{WWW}}.
\newblock


\bibitem[\protect\citeauthoryear{Zhang, Chen, Ai, Yang, and Croft}{Zhang
  et~al\mbox{.}}{2018}]%
        {zhang2018towards}
\bibfield{author}{\bibinfo{person}{Yongfeng Zhang}, \bibinfo{person}{Xu Chen},
  \bibinfo{person}{Qingyao Ai}, \bibinfo{person}{Liu Yang}, {and}
  \bibinfo{person}{W~Bruce Croft}.} \bibinfo{year}{2018}\natexlab{}.
\newblock \showarticletitle{Towards conversational search and recommendation:
  System ask, user respond}. In \bibinfo{booktitle}{\emph{CIKM}}.
  \bibinfo{pages}{177--186}.
\newblock


\end{thebibliography}

\end{document}